\documentclass[a4paper,oneside,10pt]{amsart}

\usepackage[T1]{fontenc}
\usepackage{amsmath,amssymb,epsf}
\usepackage{amsfonts,amsbsy,amscd,stmaryrd}
\usepackage{latexsym,amsopn}
\usepackage{amsthm}
\usepackage{esint}
\usepackage{mathrsfs}

\usepackage{graphicx}
\usepackage{color}
\definecolor{Blue}{rgb}{0.,0.,1.}
\definecolor{Red}{rgb}{1.,0.,0.}

\newcounter{smallarabics}
\newenvironment{arabicenumerate}
{\begin{list}{{\normalfont\textrm{(\arabic{smallarabics})}}}
  {\usecounter{smallarabics}\setlength{\itemindent}{0cm}
   \setlength{\leftmargin}{5ex}\setlength{\labelwidth}{4ex}
   \setlength{\topsep}{0.75\parsep}\setlength{\partopsep}{0ex}
   \setlength{\itemsep}{0ex}}}
{\end{list}}

\newcounter{smallroman}

\newenvironment{notations}
{\begin{list}{{\normalfont\textrm{-}}}
  {\setlength{\itemindent}{0cm}
   \setlength{\leftmargin}{2ex}\setlength{\labelwidth}{4ex}
   \setlength{\topsep}{0.75\parsep}\setlength{\partopsep}{1ex}
   \setlength{\itemsep}{1ex}}
}
{\end{list}}

\let\origmaketitle\maketitle
\def\maketitle{
  \begingroup
  \def\uppercasenonmath##1{} 
  \let\MakeUppercase\relax 
	\origmaketitle
  \endgroup
}

\newcommand{\ben}{\begin{arabicenumerate}}  
\newcommand{\een}{\end{arabicenumerate}}

\def\init{\setcounter{equation}{0}}

\newtheorem{theorem}{Theorem}[section]

\newtheorem{proposition}[theorem]{Proposition}
\newtheorem{lemma}[theorem]{Lemma}
\newtheorem{definition}[theorem]{Definition}

\newtheorem{remark}[theorem]{Remark}
\newtheorem{example}[theorem]{Example}
\newcommand{\beq}{\begin{equation}}
\newcommand{\eeq}{\end{equation}}

\newcommand{\bex}{\begin{example}}
\newcommand{\eex}{\end{example}}
\def\bel{\begin{lemma}}
\def\eel{\end{lemma}}
\def\bet{\begin{theorem}}
\def\eet{\end{theorem}}
\def\bed{\begin{definition}}
\def\eed{\end{definition}}
\def\ber{\begin{remark}}
\def\eer{\end{remark}}

\def\rr{{\mathbb R}}
\def\zz{{\mathbb Z}}
\def\cc{{\mathbb C}}
\def\nn{{\mathbb N}}

\def\bar{\overline}

\def\cinf{C^\infty}
\def\c0inf{C_0^\infty}

\def\proof{
\noindent{\bf Proof.}\ \ }

\usepackage{calrsfs}
\DeclareMathAlphabet{\pazocal}{OMS}{zplm}{m}{n}
\DeclareMathAlphabet{\mathsfsl}{OMS}{cmss}{m}{n}

\DeclareSymbolFont{altletters}  {OML}{zplm}{m}{it}
\DeclareMathSymbol{\altdelta}{\mathalpha}{altletters}{"0E}
\DeclareMathSymbol{\alteta}{\mathalpha}{altletters}{"11}

\def\cV{{\mathcal V}}

\def\cU{{\mathcal U}}

\def\cN{{\pazocal N}}

\def\cW{{\pazocal W}}

\def\wf{{\rm WF}}

\def\i{{\rm i}}

\DeclareMathOperator{\Dom}{Dom}

\def\vac{{\rm vac}}

\def\qed{$\Box$\medskip}
\newcommand{\qeds}{\qed}

\def \p{ \partial}
\def\12{\frac{1}{2}}
\def\14{\frac{1}{4}}

\def\supp{{\rm supp}}
\def\e{{\rm e}}
\newcommand{\one}{\boldsymbol{1}}
\def\cH{{\pazocal H}}

\def\coinf{C_{\rm c}^\infty}

\def\c{{\pazocal }}

\def\12{\frac{1}{2}}

\def\supp{{\rm supp}}

\def\e{{\rm e}}

\def\Diff{{\rm Diff}}
\def\rx{{\rm x}}
\def\bx{{\rm x}}

\def\bep{\begin{proposition}}
\def\eep{\end{proposition}}

\def\Op{{\rm Op}^{\rm w}}

\newcommand{\mat}[4]{\begin{pmatrix}#1 &#2  \\ #3 &#4 \end{pmatrix}}
\newcommand{\col}[2]{\begin{pmatrix}#1 \\#2\end{pmatrix}}

\def\CARal{{\rm C\hskip 0.25 em \hbox{\raise 1.72 ex 
\hbox{$\scriptscriptstyle\rm al$}\kern -0.57 em A}R}}

\def\otimesal{\mathop{\hbox{\raise 1.5 ex
  \hbox{$\scriptscriptstyle\rm al$}
\kern -0.92 em \hbox{$\otimes$}}}}
\def\oplusal{\mathop{\hbox{\raise 1.5 ex
  \hbox{$\scriptscriptstyle\rm al$}
\kern -0.92 em \hbox{$\oplus$}}}}
\def\Gammal{\hbox{\raise 1.68 ex 
\hbox{$\scriptscriptstyle\rm al$}\kern -0.50 em $\Gamma$}}
\def\Bal{\hbox{\raise 1.68 ex 
\hbox{$\scriptscriptstyle\rm  al$}\kern -0.50 em $B$}}
\def\CARal{{\rm C\hskip 0.25 em \hbox{\raise 1.72 ex 
\hbox{$\scriptscriptstyle\rm al$}\kern -0.57 em A}R}}

\DeclareMathAlphabet{\mathpzc}{OT1}{pzc}{m}{it}
\newcommand{\bra}{\langle} 
\newcommand{\ket}{\rangle}

\DeclareSymbolFont{boldoperators}{OT1}{cmr}{bx}{n}
\SetSymbolFont{boldoperators}{bold}{OT1}{cmr}{bx}{n}

\makeatletter
\newcommand*{\defeq}{\mathrel{\rlap{%
                     \raisebox{0.3ex}{$\m@th\cdot$}}%
                     \raisebox{-0.3ex}{$\m@th\cdot$}}%
                     =}
                     
\makeatother
\makeatletter
\newcommand*{\eqdef}{=\mathrel{\rlap{%
                     \raisebox{0.3ex}{$\m@th\cdot$}}%
                     \raisebox{-0.3ex}{$\m@th\cdot$}}%
                     }
\makeatother
\DeclareMathAlphabet{\mathpzc}{OT1}{pzc}{m}{it}

\def\Op{{\rm Op}}
\def\WF{{\rm WF}}

\def\altb{{\rm\textit{b}}}
\def\altc{{\rm\textit{c}}}
\def\altch{\skew3\hat{\rm\textit{c}}}
\def\altgh{\skew3\hat{\rm\textit{g}}}
\def\altgt{\skew3\tilde{\rm\textit{g}}}
\def\altk{{\rm\textit{k}}}
\def\altg{{\rm\textit{g}}}
\def\alth{{\rm\textit{h}}}
\def\altV{{\rm\textit{V}}}
\def\altW{{\rm\textit{W}}}
\def\altm{{\rm\textit{m}}}
\def\altVh{\skew5\hat{\rm\textit{V}}}
\def\altVt{\skew5\tilde{\rm\textit{V}}}
\def\altR{{\rm\textit{R}}}

\newcommand{\bea}{\begin{aligned}}
\newcommand{\beal}{\begin{array}{l}}
\newcommand{\eeal}{\end{array}}
\newcommand{\eea}{\end{aligned}}

\def\cf{C^\infty}
\def\cof{C_{\rm c}^\infty}

\def\td{{\rm td}}

\def\Htd{{\rm td}}

\def\bg{{\rm bg}}

\def\ast{\rm ast}
\def\pos{{\rm pos}}
\def\adg{{\rm ad}}
\def\dg{{\rm d}}

\def\rf{{\rm ref}}

\def\sca{{\rm out/in}}

\def\inout{{\rm in/out}}
\def\inn{{\rm in}}
\def\out{{\rm out}}

\def\spexi{{k}}

\newcommand{\traa}[1]{\mskip-6mu\upharpoonright_{#1}}
\def\pe{\overline{\p}}

\def\zero{{\mskip-4mu{\rm\textit{o}}}}

\def\ry{{\rm y}}
\def\cinfb{\cinf_{\rm b}}
\def\BT{{\rm BT}}

\def\outin{{\rm out/in}}

\def\varT{t_{+}}

\def\sobo{{m}}
\def\AH{H}

\begin{document}
\title[Hadamard property of the \emph{in} and \emph{out} states on asymptotically static spacetimes]{\large Hadamard property of the \emph{in} and \emph{out} states \\ for Klein-Gordon fields on asymptotically static spacetimes}
\author{}
\address{Universit\'e Paris-Sud XI, D\'epartement de Math\'ematiques, 91405 Orsay Cedex, France}
\email{christian.gerard@math.u-psud.fr}
\author{\normalsize Christian \textsc{G\'erard} \& Micha{\l} \textsc{Wrochna}}
\address{Universit\'e Grenoble Alpes, Institut Fourier, UMR 5582 CNRS, CS 40700, 38058 Grenoble \textsc{Cedex} 09, France}
\email{michal.wrochna@univ-grenoble-alpes.fr}
\begin{abstract}We consider the massive Klein-Gordon equation on a class of asymptotically static spacetimes (in the long range sense) with Cauchy surface of bounded geometry. We prove the existence and Hadamard property of the \emph{in} and \emph{out} states constructed by scattering theory methods.
\end{abstract}

\maketitle

\section{Introduction \& summary}

\subsection{Hadamard property of $\rm in/out$ states}
The construction of quantum states from scattering data is a subject that has been studied extensively in various contexts in Quantum Field Theory.  Let us mention for example the  wave and Klein-Gordon fields  on Minkowski space, in external electromagnetic potentials \cite{isozaki,lundberg,rui,seiler}, or on curved spacetimes with special a\-symp\-totic symmetries, \cite{wald0,DK0,DK1,DK2,Mo1}. On the physics side, the primary motivation is to give meaning to the notion of particles and anti-particles and to describe quantum scattering phenomena. 

From the mathematical point of view, the problems often discussed in this context in the literature involve existence of scattering and M{\o}ller operators, the question of asymptotic completeness, or specific properties of states such as the ground state or thermal condition with respect to an asymptotic dynamics, see e.g. \cite{drouot,DD,DRS,GGH,guillaume,nicolas} for various recent developments on curved backgrounds.

In the present paper we address the question of whether the so-called \textit{in} and \textit{out} states on asymptotically static spacetimes satisfy the \emph{Hadamard condition} \cite{KW}. Nowadays regarded as an indispensable ingredient in the perturbative construction of interacting fields (see e.g. recent reviews \cite{HW,KM,FV2}), this property accounts for the correct short-distance behaviour of two-point functions. It can be conveniently formulated as a condition on the \emph{wave front set} of the state's two-point functions \cite{radzikowski} --- a terminology that we explain in the paragraphs below. It is known that in the special case of the conformal wave equation, one can study the wave front set of the two-point functions quite directly in the geometrical setup of conformal scattering on asymptotically flat spacetimes \cite{Mo2,characteristic} (cf. \cite{DMP1,DMP2,BJ} for generalizations on the allowed classes of spacetimes). Furthermore, propagation estimates in ${\rm b}$-Sobolev spaces of variable order were used recently to show a similar result in the case of the wave equation on asymptotically Minkowski spacetimes \cite{VW}, drawing on earlier developments by Vasy et al. \cite{BVW,semilinear,kerrds,positive}. The two  methods being however currently limited to a special value of the mass parameter, our focus here is instead on the proof of the Hadamard property of the \textit{in} and \textit{out} state for the Klein-Gordon operator $P=-\Box_\altg+\altm^2$ for any positive mass $\altm$, or more generally for $P=-\Box_\altg+\altV$ with a real-valued potential $\altV\in\cf(M)$ satisfying an asymptotic  positivity condition. \medskip

\subsection{The model problem}
We first consider the special case of a $1+d$-dimensional globally hyperbolic spacetime $(M,\altg)$ with Cauchy surface $\Sigma$ and metric of the form $\altg=-dt^2 + \alth_t$, with $\alth_t$ a Riemannian metric smoothly depending on $t$. The Klein-Gordon operator can be written in the form
\beq\label{Pintro}
P=\p_t^2 + r(t)\p_t + a(t,\bx,\p_\bx),
\eeq
where $r(t)$ is the multiplication operator $|\alth_t|^{-\12}\p_t |\alth_t|^{\12}$ and $a(t,\bx,D_\bx)\in\Diff^2(\Sigma)$ has principal symbol  $\spexi \cdot \alth_t^{-1}(\bx)\spexi$ (where $\xi=(\tau,k)$ is the dual variable of $x=(t,\bx)$). Now, supposing $\Sigma$ is a \emph{manifold of bounded geometry} (see Subsect. \ref{ss:mbg}), there exist uniform pseudo\-differential operator classes $\Psi^m(\Sigma)$ due to Kordyukov and Shubin \cite{Ko,Sh2} that generalize the well-known pseudo\-differential calculus of H\"ormander on $\rr^d$ and closed manifolds. Here in addition, in order to control decay in time, we introduce $t$-dependent pseudo\-differential operators $\Psi^{m,\delta}_\td(\rr;\Sigma)$ as quantizations of $t$-dependent symbols $a(t, \rx, \spexi)$ that satisfy
\[
|\p^{\alpha}_{t}\p_{\rx}^{\beta}\p_{\spexi}^{\gamma}a(t, \rx, \spexi)|\leq C_{\alpha\beta\gamma}\langle t\rangle^{\delta- \alpha}\langle \spexi\rangle^{m- |\gamma|}, \ \ \alpha\in \nn, \ \beta, \gamma\in \nn^{d},
\]
where $\bra t\ket=(1+t^2)^{\12}$, $\bra \spexi\ket=(1+|\spexi|^2)^{\12}$, and the constants $C_{\alpha\beta\gamma}$ are uniform in an appropriate sense. This allows us to state a hypothesis that accounts for asymptotic ultra-staticity of $(M,\altg)$ at future and past infinity. Namely, we assume that there exists $\delta>0$ and $a_\out,a_\inn\in\Psi^2(\Sigma)$ elliptic such that on $\rr^{\pm}\times\Sigma$,
\[
(\Htd) \  \ \ \beal
a(t, \rx, D_{\rx})- a_{\outin}(\rx, D_{\rx})\in \Psi_{\td}^{2, -\delta}(\rr; \Sigma),\\[2mm]
r(t)\in \Psi_{\td}^{0, -1-\delta}(\rr; \Sigma).
\eeal  
\]
In practice, in our main cases of interest $a_{\outin}(\rx, D_{\rx})$ will simply be the Laplace-Beltrami operator of some asymptotic metric $\alth_\outin$ plus the mass or potential term. 
 
On top of that, an important condition that we assume is the {\em strict positivity} of $a_{\outin}(\rx, D_{\rx})$, namely:
\[
a_{\outin}(\rx, D_{\rx})\geq m^{2}>0.
\]
This condition has two consequences. First of all, it ensures that the vacuum states $\omega_{\outin}^{\rm vac}$ for the Klein-Gordon operators $P_{\outin}= \p_{t}^{2}+ a_{\outin}(\rx, , D_{\rx})$ have covariances given by pseudodifferential operators (in the uniform classes $\Psi^m(\Sigma)$). Secondly, it allows to control the differences  of fractional powers $a(t, \rx, D_{\rx})^{\alpha}- a_{\outin}(\rx, D_{\rx})^{\alpha}$ when $t\to \infty$, see Prop. \ref{l5.1}. This control is an important technical tool in  Sect. \ref{sec2}.

Let now $\cU(t,s)$ be the Cauchy evolution of $P$, i.e. the operator that maps Cauchy data of $P$ at time $s$ to Cauchy data at time $t$. 
In this setup, what we call \emph{time-$t$ covariances of the $\out$ state} are the pair of operators defined by
\beq\label{eq:thelimit}
c_{\out}^\pm(t)\defeq \lim_{t_{+}\to+\infty} \cU(t,t_{+}) c_\out^{\pm,\vac}  \cU(t_{+},t)
\eeq
whenever the limit exists (in a sense made precise later on), where $c_\out^{\pm,\vac}$ equals
\[
c_\out^{\pm,\vac}= \12 \begin{pmatrix}\one & \pm a_\out^{-\12} \\ \pm a_\out^{\12} & \one\end{pmatrix}.
\]
To elucidate the interpretation of $c_\out^{\pm,\vac}$ let us point out that $c_\out^{\pm,\vac}$ is the spectral projection on $\rr^{\pm}$ of the generator\footnote{This generator is selfadjoint for the energy scalar product.} of the Cauchy evolution $\cU_\out(t,s)$ corresponding to the asymptotic Klein-Gordon operator $P_\out\defeq\p_t^2+a_\out$. On the other hand, to $c_\out^{\pm,\vac}$, $c_\out^\pm$ we can associate pairs of operators $\Lambda_\out^{\pm,\vac}$, $\Lambda_\out^\pm: \cf_{\rm c}(M)\to \cf(M)$ by
\[
\bea
\Lambda_\out^{\pm,\vac}(t,s)&\defeq \mp\pi_0\cU_\out(t,0)c_{\out}^{\pm,\vac}\cU_\out(0,s)\pi_1^*, \\ \Lambda_\out^\pm(t,s)&\defeq \mp\pi_0\cU(t,0)c_{\out}^\pm(0)\cU(0,s)\pi_1^*,
\eea
\]
where we wrote $\Lambda_\out^{\pm,\vac}$, $\Lambda_\out^\pm$ as operator-valued Schwartz kernels in the time variable and $\pi_{i}: \coinf(\Sigma)^{2}\ni\col{f_{0}}{f_{1}}\mapsto f_{i}\in \coinf(\Sigma)$ are the projections to the two components  of Cauchy data. 

In QFT terms (strictly speaking, using the terminology for charged fields), the operators $\Lambda_\out^{\pm,\vac}$, $\Lambda_\out^\pm$ are \emph{two-point functions}, i.e. they satisfy
\[
\bea
P_\out \Lambda^{\pm,\vac}_{\out}=\Lambda^{\pm,\vac}_\out P_\out = 0,& \ \ \Lambda^{+,\vac}_\out-\Lambda^{-,\vac}_\out=\i G_\out, \ \ \Lambda^{\pm,\vac}_{\out}\geq 0, \\
P \Lambda^\pm_{\out}=\Lambda^\pm_{\out} P = 0, \quad & \ \ \Lambda^+_{\out}-\Lambda^-_{\out}=\i G, \quad \Lambda^\pm_{\out}\geq 0,
\eea
\] 
where $G_\out$, $G$ are the \emph{causal propagators} for respectively $P_\out$, $P$, i.e. 
\[
G_\out(t,s)=\i \pi_0\cU_\out(t,s)\pi_1^*, \ \ G(t,s)=\i \pi_0\cU(t,s)\pi_1^*.
\] 
As a consequence, using the standard apparatus of algebraic QFT one can associate states $\omega_\out^\vac$, $\omega_\out$ on the corresponding ${\rm CCR}$ $C^*$-algebras: $\omega_\out^\vac$ is then the very well studied ground state associated with $P_\out$ and $\omega_\out$ is the \emph{out} state that we study.

Our first result can be expressed as follows in terms of the two-point functions $\Lambda^\pm_\out$. 

\begin{theorem}\label{thm:main1} Assume $(\Htd)$. Then the limit \eqref{eq:thelimit} exists and $\omega_\out$ is a Hadamard state, i.e.
the two-point functions $\Lambda^\pm_\out$ satisfy the {Hadamard condition}:
\beq\label{eq:had1}
\wf'(\Lambda^\pm_\out)\subset  \cN^\pm\times\cN^\pm,
\eeq
where $\cN^+$, $\cN^-$ are the two connected components of the characteristic set $\cN\subset T^*M\setminus\,\zero$ of $P$.
\end{theorem}

Above, $\wf'(\Lambda_\out^\pm)$ stands for the primed wave front set of $\Lambda_\out^\pm$, i.e. it is the image of the wave front set of the (full) Schwartz kernel of $\Lambda_\out^\pm$ by the map $(x,\xi,x',\xi')\mapsto (x,\xi,x',-\xi')$. We recall that the wave front set of a distribution characterizes the location $(x,x')$ of its singularities, as well as the responsible directions $(\xi,\xi')$ in Fourier space, see \cite{hoermander} for the precise definition and basic properties, cf. \cite{BDH} for a concise introduction. The \emph{characteristic set} of $P$ is by definition $\cN=p^{-1}(\{0\})$  understood as a subset of $T^*M\setminus\zero$ (where $\,\zero$ is the zero section of the cotangent bundle), where $p(x,\xi)=\xi\cdot \altg^{-1}(x)\xi$ is the principal symbol of $P$.  


The essential feature of the Hadamard condition \eqref{eq:had1} is that it constraints $\wf'(\Lambda^\pm_\out)$ to the positive/negative frequency components $\cN^\pm\times\cN^\pm$ (rather than merely to $\cN\times\cN$, as would be the case for very general classes of bi-solutions). Thus, on a very heuristic level, the plausibility of this statement can be explained as follows. In a static situation, $c^{\pm,\vac}_\out$ are interpreted as projections that single out Cauchy data that propagate as superpositions of plane waves with positive/negative frequency, and thus with wave front set in $\cN^\pm$. On a generic asymptotically flat spacetime it is not immediately clear what the analogous decomposition at finite times is, but instead one can try to use the decomposition given by $c^{\pm,\vac}_\out$ at \emph{infinite times}: this is what indeed motivates the definition of $\Lambda^\pm_{\out}$. The crucial difficulty is however to control the wave front set of the infinite time limit \eqref{eq:thelimit}. \medskip

In addition to the statement of Thm. \ref{thm:main1}, we get in a similar vein a Hadamard state $\omega_\inn$ by taking the limit analogous to \eqref{eq:thelimit} with $t_- \to -\infty$ instead of $t_+\to + \infty$ and   $c^{\pm,{\rm vac}}_{\rm in}$ instead of $c^{\pm, {\rm vac}}_{\rm out}$. This is the so-called \emph{in} state.

\subsection{General asymptotically static spacetimes }
Our results extend to a more general class of asymptotically static spacetimes $M=\rr\times\Sigma$ with metric of the form
\[
\altg= - \altc^{2}(x)dt^{2}+ (d\rx^{i}+ \altb^{i}(x)dt)\alth_{ij}(x)(d\rx^{j}+ \altb^{j}(x)dt),
\]
where $(\Sigma,\alth)$ is a manifold of bounded geometry and $\altc,\alth,\altb$ as well as their inverses are bounded with all derivatives (with respect to the norm defined using a reference Riemannian metric). By \emph{asymptotically static} we mean that there exist Riemannian metrics  $\alth_\outin$ and smooth functions $\altc_{\outin}$ on $\Sigma$, such that on $\rr^\pm\times\Sigma$,
\[
(\ast)\ \ \beal
\alth(x)- \alth_{\outin}(\rx)\in S^{-\mu},\\[2mm]
\altb(x)\in S^{-\mu'},  \mbox{ and } \altc(x)- \altc_{\outin}(\rx)\in S^{-\mu}
\eeal
\]
for some $\mu>0$, $\mu'>1$; in a similar vein the potential $\altV$ is required to satisfy $\altV(x)- \altV_{\outin}(\rx)\in S^{-\mu}$ for some smooth $\altV_{\outin}$. Above, the notation $f\in S^{-\mu}$ means symbolic decay in time, i.e. $\p^{\alpha}_{t}f\in O(\langle t\rangle^{-\mu- |\alpha|})$ for all $\alpha\in \nn^{1+d}$; we refer to Subsect. \ref{ss:asast} for the precise formulation.

In this more general situation, the Klein-Gordon operator is not necessarily of the form \eqref{Pintro} considered so far. However, under a positivity assumption $(\pos)$ on $\altV_{\outin}$, it turns out that there are natural coordinates in terms of which the Klein-Gordon operator is very closely related to an operator \eqref{Pintro} satisfying $(\Htd)$, i.e. one is obtained from the other by conjugation with some multiplication operators. This allows us to give a very similar definition of the \textit{out}/\textit{in} state $\omega_\outin$ and to prove a direct analogue of Thm. \ref{thm:main1}.  

\subsection{Outline of proofs} The importance of our result stems from the fact that it brings together for the first time  methods from scattering theory and the analysis of Hadamard states. The main technical ingredient that we use in the proof of our theorem is an approximate diagonalization\footnote{On a side note, let us mention that a different diagonalization procedure was proposed by Ruzhansky and Wirth in the context of dispersive estimates \cite{RW,wirth}; in their method it is the (full) symbol of the generator of the Cauchy evolution that is diagonalized (rather than the Cauchy evolution itself).} of the Cauchy evolution by means of elliptic pseudo\-differential operators, derived in detail in \cite{bounded} and based on the strategy developed successively in the papers \cite{junker,JS,GW,GW2}. Specifically, its outcome is that the Cauchy evolution of $P$ can be written as
\beq\label{eq:modsmoj}
\cU(s,t)=T(t) \cU^{\adg}(t,s) T(s)^{-1}
\eeq
where $T(t)$ is a $2\times 2$ matrix of pseudo\-differential operators (smoothly depending on $t$) and $\cU^{\adg}(t,s)$ is `almost diagonal' (hence the superscript $\adg$). Namely, $\cU^{\adg}(s,t)$ is the Cauchy evolution of a time-dependent operator of the form $\i\p_t+H^{\adg}(t)$, where
\[
H^{\adg}(t)=\mat{\epsilon^{+}(t)}{0}{0}{\epsilon^-(t)}
\]
modulo smooth terms (more precisely, modulo terms in $C^\infty(\rr^2,\cW^{-\infty}(\Sigma)\otimes \cc^2)$, where $\cW^{-\infty}(\Sigma)$ are the operators that map $H^{-m}(\Sigma)$ to $H^{m}(\Sigma)$ for each $m\in\nn$), and $\epsilon^\pm(t)$ are elliptic pseudodifferential operators of order $1$ with principal symbol $\pm(k\cdot \alth_t^{-1} k)^{\12}$. Now, because of this particular form of the principal symbol, solutions of $(\i^{-1}\p_t +\epsilon^\pm(t))$ propagate with wave front set in $\cN^\pm$. This allows us  to prove that if we fix some $t_0\in\rr$ and set
\[
c^\pm_{\rm ref}(t_0)\defeq T(t_0)\pi^\pm  T^{-1}(t_0), \ \mbox{ where \ } \pi^+=\mat{\one}{0}{0}{0}, \ \ \pi^-=\mat{0}{0}{0}{\one},
\]
then $\Lambda_{\rm ref}^\pm(t,s)\defeq \mp\pi_0\cU(t,t_0)c_{\rm ref}^\pm(t_0)\cU(t_0,s)\pi_1^*$ have wave front set only in $\cN^\pm\times\cN^\pm$ and therefore satisfy the Hadamard condition \eqref{eq:had1}. As a consequence, to prove the Hadamard condition for $\Lambda^\pm_{\inout}$ it suffices to show that 
\[
c^\pm_\inout - c_{\rm ref}^\pm \in \cW^{-\infty}(\Sigma)\otimes B(\cc^2).
\]
To demonstrate that this is the case, we use assumption $(\Htd)$ to control the decay in time of various remainders in identities `modulo smooth'. The most crucial estimate here is
\beq\label{eq:tima}
H^{\adg}(t)-\begin{pmatrix}a(t)^{\12} & 0 \\ 0 & -a(t)^{\12} \end{pmatrix}\in \Psi_\td^{0,-1-\delta}(\rr;\Sigma)\otimes B(\cc^2)
\eeq
for large $t$, which then yields time-decay of various commutators that appear in the proofs. We obtain \eqref{eq:tima} by revisiting the approximate diagonalization \eqref{eq:modsmoj} using poly-homogeneous expansions of pseudodifferential operators in $\Psi_\td^{m,-\delta}(\rr;\Sigma)$ in both $m$ and $\delta$; more details are given in Sect. \ref{secscat}. This requires us to study the classes $\Psi_\td^{m,-\delta}(\rr;\Sigma)$ carefully, in particular we employ a variant of Seeley's theorem on powers of pseudo\-differential operators elliptic in the standard $\Psi^m$ sense.

\subsection{Plan of the paper} The paper is structured as follows.

In Sect. \ref{sec1} we fix some basic terminology and recall the definition of two-point functions and covariances of states in the context of non-interacting Quantum Field Theory. 

Sect. \ref{secbounded} contains a brief overview of the pseudo\-differential calculus on manifolds of bounded geometry.  We then introduce the time-dependent pseudodifferential operator classes $\Psi_\td^{m,\delta}$ and study some of their properties. 

In Sect. \ref{sec2} we recall the approximate diagonalization of the Cauchy evolution used in \cite{bounded} to construct generic Hadamard states. We then give a refinement in the setup of assumption $(\Htd)$ by showing time decay of various remainder terms.

Sect. \ref{inout} contains the construction of the \textit{in} and \textit{out} states and the proof of their Hadamard property in the case of asymptotically static spacetimes (assumptions $(\ast)$ and $(\pos)$). The key ingredients are the reduction to the setup of assumption $(\Htd)$ and the estimates obtained in Sect. \ref{sec2}.

Various auxiliary proofs are collected in Appendix \ref{secapp1}. 

\section{Preliminaries}\init\label{sec1}

\subsection{Notation}\label{sec1.1}The space of differential operators (of order $m$) over a smooth manifold $M$ (here always without boundary) is denoted  $\Diff(M)$ ($\Diff^m(M)$). The space of smooth functions on $M$ with compact support is denoted $\coinf(M)$.

The operator of multiplication by a function $f$ will be denoted by $f$, while the operators of partial differentiation will be denoted by $\pe_{i}$, so that $[\pe_{i}, f]= \p_{i}f$.

\begin{notations}
\item If $a, b$ are selfadjoint operators on a Hilbert space $\cH$, we write $a\sim b$ if
\[
  a,b >0, \ \ \Dom a^{\12}= \Dom b^{\12}, \ \ c^{-1}b\leq a \leq cb,
\]
for some  constant $c>0$.

\item Similarly, if  $I\subset \rr$ is an open interval and $\{\cH_{t}\}_{t\in I}$ is a family of Hilbert spaces with $\cH_{t}= \cH$ as topological vector spaces, and $a(t), b(t)$ are two selfadjoint operators on   $\cH_{t}$, we write $a(t)\sim b(t)$ if for each $J\Subset I$ there exist constants $c_{1, J}, c_{2, J}>0$  such that
  \beq\label{equitd}
  a(t), b(t) \geq c_{1, J}>0, \ \  c_{2, J}b(t)\leq a(t) \leq c_{2, J}^{-1}b(t), \ t\in J.
\eeq
\end{notations}

\subsection{Klein-Gordon operator}\label{ssec:classical}

Let $(M,\altg)$ be a Lorentzian spacetime (we use the convention $(-,+,\dots,+)$ for the Lorentzian signature).  We consider the Klein-Gordon operator with a real-valued potential $\altV\in\cf(M)$ 
\[
P=-\Box_g + \altV \in\Diff^2(M), 
\]
Since $\altV$ is real-valued we have $P=P^*$ in the sense of formal adjoints with respect to the $L^2(M,g)$ scalar product, canonically defined using the volume form.  

For $K\subset M$ we denote $J_\pm(K)\subset M$ its causal future/past, see e.g. \cite{BF,W}. Let $C^\infty_\pm(M)$ be the space of smooth functions whose support is future or past compact, that is
\[
C^\infty_\pm(M) = \{ f\in \cf(M) : \  \supp f\subset J_\pm(K) \mbox{ \ for\ some\ compact\ } K\subset M\}.
\]
We assume that $(M,\altg)$ is globally hyperbolic, i.e. admits a foliation by Cauchy surfaces\footnote{Let us recall that a Cauchy surface is a smooth hypersurface that is intersected by every inextensible, non-spacelike (i.e. causal) curve exactly once.} (in the next sections we will impose more restrictive conditions on $(M,\altg)$, but these are irrelevant for the moment). It is well known that $P$ has then unique \emph{advanced}/\emph{retarded propagators}, i.e. operators $G_\pm:C^\infty_\pm(M)\to C^\infty_\pm(M)$ s.t.
\beq\label{eq:Pinverse}
 P G_\pm = \one \mbox{ \ on \ } C^\infty_\pm(M).
\eeq
A standard duality argument that uses $P=P^*$, (\ref{eq:Pinverse}), and the fact that $C^\infty_+(M)\cap C^\infty_-(M)=\cof(M)$ on globally hyperbolic spacetimes,  gives $G_+^*=G_-$ as sesquilinear forms on $\cof(M)$. The \emph{causal propagator} (often also called \emph{Pauli-Jordan commutator function}) of $P$ is by definition $G\defeq G_+-G_-$, interpreted here as a map from  $\cof(M)$ to  $C^\infty_+(M)+C^\infty_-(M)$, the space of space-compact smooth functions.

\subsection{Symplectic space of solutions} In what follows we recall the relation between quasi-free states, two-point functions, and field quantization. The reader interested only in the analytical aspects can skip this discussion and move directly to equations \eqref{enlambda1}--\eqref{enlambda3}, which can be taken as the definition of two-point functions in the present context. 

By a \emph{phase space} we will mean a pair $(\cV,q)$ consisting of a complex vector space $\cV$ and a non degenerate hermitian form $q$ on $\cV$. In our case the phase space of interest (i.e. the phase space of the classical non-interacting scalar field theory) is
\beq\label{defo}
\cV\defeq \frac{\cof(M)}{P\cof(M)}, \quad \bar u \,q v\defeq \i^{-1}(u| G v),
\eeq
where $(\cdot|\cdot)$ is the $L^2(M,\altg)$ pairing. The sesquilinear form $q$ is indeed well-defined on the quotient space $\cof(M)/P\cof(M)$ because $PG=GP=0$ on test functions. Using that $G_+^*=G_-$ one shows that $q$ is hermitian, and it is also not difficult to show that it is non-degenerate.

 Note that in contrast to most of the literature, we work with hermitian forms rather than with real symplectic ones, but the two approaches are equivalent.

\subsection{States and their two-point functions}\label{ss:qfree}

Let $\cV$ be a complex vector space, $\cV^{*}$ its anti-dual and $L_{\rm h}(\cV, \cV^{*})$ the space of hermitian sesquilinear forms on $\cV$. If $q\in L_{\rm h}(\cV, \cV^{*})$ then we can define the { polynomial  CCR $*$-algebra ${\rm CCR}^{\rm pol}(\cV,q)$ (see e.g. \cite[Sect. 8.3.1]{derger}) \footnote{See also \cite{GW,wrothesis} for remarks on the transition between real and complex vector space terminology.}. It is constructed as the span of the so-called {\em abstract complex fields} $\cV\ni v\mapsto \psi(v), \psi^{*}(v)$, which are taken to be anti-linear, resp. linear in $v$ and are subject to  the canonical commutation relations
\[
[\psi(v), \psi(w)]= [\psi^{*}(v), \psi^{*}(w)]=0,  \ \ [\psi(v), \psi^{*}(w)]=  \bar{v} q w \one, \ \ v, w\in \cV.
\]
Our main object of interests are the states\footnote{Let us recall that a state $\omega$ is a linear functional on ${\rm CCR}^{\rm pol}(\cV,q)$ such that $\omega(a^* a)\geq 0$ for all $a$ in ${\rm CCR}^{\rm pol}(\cV,q)$, and $\omega(\one)=1$.} on ${\rm CCR}^{\rm pol}(\cV,q)$.

The \emph{complex covariances}  $\Lambda^\pm\in L_{\rm h}(\cV,\cV^*)$ of a state $\omega$ on ${\rm CCR}^{\rm pol}(\cV,q)$ are defined in terms of the abstract field operators by
\beq\label{eq:lambda}
\bar{v}\Lambda^+ w = \omega\big(\psi(v)\psi^*(w)\big), \quad \bar{v}\Lambda^- w = \omega\big(\psi^*(w)\psi(v)\big), \quad v,w\in \cV
\eeq
Note that both $\Lambda^\pm$ are positive and by the canonical commutation relations one always has $\Lambda^+ - \Lambda^- = q$. Conversely, if one has a pair of hermitian forms $\Lambda^\pm$ such that  $\Lambda^+ - \Lambda^- = q$ and $\Lambda^\pm\geq 0$ then there is a unique \emph{quasi-free }state $\omega$ such that (\ref{eq:lambda}) holds. 
We will thus further restrict our attention to quasi-free states and more specifically to their complex covariances $\Lambda^\pm$. 

In QFT (at least for scalar fields) the phase space of interest is the one  defined in (\ref{defo}). In that specific case it is convenient to consider instead of complex covariances a pair of operators $\Lambda^\pm:\cof(M)\to\cf(M)$ such that 
\beq\label{eq:lambdab}
(v|\Lambda^+ w) = \omega\big(\psi(v)\psi^*(w)\big), \quad (v|\Lambda^- w) = \omega\big(\psi^*(w)\psi(v)\big), \quad v,w\in \cof(M).
\eeq
We call $\Lambda^\pm$ the \emph{two-point functions} of the state $\omega$ and identify them with the associated complex covariances whenever possible. Note that because  $(\cdot|\Lambda^\pm\cdot)$ has to induce a hermitian form on the quotient space $\cof(M)/P \cof(M)$, the two-point functions have to satisfy $P\Lambda^\pm=\Lambda^\pm P=0$ on $\cof(M)$. By the Schwartz kernel theorem we can further identify $\Lambda^\pm$ with a pair of distributions on $M\times M$, these are then bi-solutions of the Klein-Gordon equation. 

In QFT on curved spacetime one is especially interested in the subclass of quasi-free \emph{Hadamard states} \cite{KW,radzikowski}. These can be defined as in the introduction, i.e. by requiring that the primed wave front set of the  Schwartz kernel of $\Lambda^\pm$ is contained in $\cN^\pm\times \cN^\pm$ (cf. e.g. \cite{radzikowski,SV,sanders} for a discussion of various equivalent formulations), $\cN^\pm\subset T^*M\setminus\zero$ being the two connected components of the characteristic set of $P$ (and $\zero\subset T^*M$ the zero section). To sum this up, specifying a Hadamard state amounts to constructing a pair of operators $\Lambda^\pm:\cof(M)\to\cf(M)$ satisfying the properties:
\begin{flalign}
\label{enlambda1} &P\Lambda^\pm=\Lambda^\pm P=0, \ \ \Lambda^+-\Lambda^-=\i G,\\
\label{enlambda2} &\Lambda^\pm\geq 0,\\
\label{enlambda3} &\wf'(\Lambda^\pm)\subset \cN^\pm\times\cN^\pm.
\end{flalign}
Existence of two-point functions as above was proved in \cite{FNW}, and an alternative argument was given in \cite{GW}, followed by the construction of a very large class of examples in \cite{bounded}. The importance of Hadamard states is primarily due to their pivotal role in renormalization on curved spacetimes \cite{BF00,HW1,HW2,dang}, see \cite{FV2,KM,HW} for recent reviews.  

Here we will be interested in showing \eqref{enlambda3} for specific two-point functions with prescribed asymptotic properties, motivated by the conceptual need for \emph{distinguished} Hadamard states whenever allowed by the spacetime geometry.

\subsection{Cauchy data of two-point functions}\label{ssec:cauchy} We will need a version of two-point functions acting on Cauchy data of $P$ instead of spacetime quantities such as $\Lambda^\pm$. To this end, let $\{ \Sigma_s\}_{s\in \rr}$ be a foliation of $M$ by Cauchy surfaces (since all $\Sigma_s$ are diffeomorphic we occasionally write $\Sigma$ instead). We define the map
\[
\varrho_s   u \defeq (u, \i^{-1} n^a\nabla_a u)\traa{\Sigma_s},
\]
acting on distributions $u$ such that the restriction ${}\traa{\Sigma_s}$ makes sense, where $n^a$ is the unit normal vector to $\Sigma_s$. It is well-known that $\varrho_s\circ G$ maps $\cof(M)$ to $\cof(\Sigma_{s})$ and that there exists an operator $G(s)$ acting on $\cof(\Sigma)\otimes\cc^2$ (not to be confused with $G$) that satisfies
\beq\label{eq:idGs}
G \eqdef (\varrho_s G)^* \circ  G(s) \circ \varrho_s G,
\eeq
where $(\varrho_s G)^*$ is the formal adjoint of $\varrho_s \circ G$ wrt. the $L^2$ inner product on $\Sigma_s\sqcup\Sigma_s$ respective to some density (that can depend on $s$, later on we will make that choice more specific).  We also set
\[
q(s)\defeq \i^{-1} G(s),
\]
so that $q(s)^*=q(s)$. 

The next result provides a Cauchy surface analogue of the two-point functions $\Lambda^\pm$, cf.  \cite{GW2} for the proof.
\begin{proposition}\label{minusu}
For any $s\in\rr$ the maps:
 \beq\label{eq:nimnim}\lambda_{}^{\pm}(s)\mapsto \Lambda^{\pm}\defeq  (\varrho_s G)^{*}\lambda^{\pm}(s)(\varrho_s G),
 \eeq
 and 
 \begin{equation}
\label{eq:nim}
\Lambda^{\pm}\mapsto\lambda^{\pm}(s)\defeq (\varrho^{*}_s G(s))^{*} \Lambda^{\pm} (\varrho^{*}_s G(s))
\end{equation}
are bijective and inverse from one another.
\end{proposition}
It is actually convenient to make one more definition and set:
\beq\label{eq:nimnimnim}
c^\pm(s)=\pm \i^{-1} G(s) \lambda^\pm(s) : \cof(\Sigma)\otimes B(\cc^2)\to \cf(\Sigma)\otimes B(\cc^2).
\eeq
We will simply call $c^\pm(s)$ the \emph{(time-$s$) covariances of the state $\omega$}. A pair of operators $c^\pm(s)$ are covariances of a state iff the operators $\Lambda^\pm$ defined by (\ref{eq:nimnim}) and (\ref{eq:nimnimnim}) satisfy (\ref{enlambda1})-(\ref{enlambda2}), which is equivalent to the conditions
\begin{flalign}
&c^+(s)+c^-(s)=\one,\label{eq:secondcondlam0}\\
&\lambda^\pm(s)\geq 0,\label{eq:secondcondlam}
\end{flalign}
where we identified the operators $\lambda^\pm(s)$ with hermitian forms using the same pairing as when we took the formal adjoint in \eqref{eq:idGs}. Note that \eqref{eq:secondcondlam0} can also be expressed as $\lambda^+(s)-\lambda^-(s)=q(s)$. 
 
Additionally, a state (recall that we consider only quasi-free states) is \emph{pure} iff its covariances $c^\pm(s)$ extend to projections on the completion of $\cof(\Sigma)\otimes\cc^2$ w.r.t. the inner product given by $\lambda^+ + \lambda^-$. In practice it is sufficient to construct $c^\pm(s)$ as projections acting on a space that is big enough to contain $\cof(\Sigma)\otimes\cc^2$, but small enough to be contained in the Hilbert space associated to $\lambda^+ + \lambda^-$.
 
\subsection{Propagators for the Cauchy evolution}\label{ssec:propaCauchy}

Recall that we have defined the operator $G(s)$ via the identity
\beq\label{eq:idGs2}
G \eqdef (\varrho_s G)^* \circ  G(s) \circ \varrho_s G.
\eeq
A direct consequence is that the operator $G^* \varrho_s G(s)$ assigns to Cauchy data on $\Sigma_s$ the corresponding solution. Similarly, for $t,s\in\rr$ the operator
\beq\label{eq:defce}
\cU(s,t)\defeq \varrho_s G^* \varrho_t^* G(t)
\eeq
produces Cauchy data of a solution on $\Sigma_s$ given Cauchy data on $\Sigma_t$. We will call $\{ \cU(s,t)\}_{s,t\in\rr}$ the \emph{Cauchy evolution} of $P$. A straightforward computation gives the group property
\beq
\cU(t,t)=\one, \quad \cU(s,t')\cU(t',t)=\cU(s,t), \ \ t'\in\rr;
\eeq
and the conservation of the symplectic form by the evolution
\beq\label{eq:conserveq}
\cU^*(s,t)q(s)\cU(s,t)=q(t).
\eeq
These identities allow to conclude that the covariances $c^\pm(t)$ (and two-point functions $\lambda^\pm(t)$) at different `times' of a quasi-free state are related by
\beq\label{eq:tiatc}
\bea
\lambda^\pm(t)&=\cU(s,t)^* \lambda^\pm(s) \cU(s,t),\\
c^\pm(t) &=\cU(t,s) c^\pm(s) \cU(s,t).
\eea
\eeq
Notice that this induces a splitting of the evolution in two parts:
\[
\cU(s,t)=\cU^+(s,t)+\cU^-(s,t), \mbox{ \ with \ } \cU^\pm(s,t)=\cU(s,t)c^\pm(t).
\]
If the state is pure then $c^\pm(t)$ are projections for all $t$ and the operators $\cU^\pm(s,t)$ obey the composition formula
\[
\cU^\pm(s,t')\cU^\pm(t',t)=\cU^\pm(s,t), \ \ \cU^\pm(s,t')\cU^\mp(t',t)=0, \ \ t'\in\rr.
\]
Let us stress that $\cU^\pm(t,t)$ is not the identity, but rather equals $c^\pm(t)$. Furthermore, if the state is Hadamard then $\cU(s,t)c^\pm(t)$ propagate singularities along $\cN^\pm$ (see the discussion in \cite{GW2}). In Sect. \ref{sec2} we will be interested in the reversed argument, namely we will construct covariances $c^\pm(t)$ of pure Hadamard states from a splitting of the evolution $\cU(t,s)$ into two parts that propagate singularities along respectively $\cN^+$, $\cN^-$.  

\section{Pseudodifferential calculus on manifolds of bounded geometry}\init\label{secbounded}
\subsection{Manifolds of bounded geometry}\label{ss:mbg} In the present section we introduce manifolds of bounded geometry and review the pseudodifferential calculus of Kordyukov and Shubin \cite{Ko,Sh2}, making also use of some results from \cite{bounded}.

Let us denote by $\altdelta$ the flat metric on $\rr^{d}$ and by $B_{d}(\ry, r)\subset \rr^{d}$ the open ball of center $\ry$ and radius $r$.


 

If $(\Sigma, \alth)$ is a $d-$dimensional Riemannian manifold and $X$ is a  $(p,q)$ tensor  on $\Sigma$, we can define the canonical norm of $X(\rx)$, $\rx\in \Sigma$, denoted by $\|X\|_{\rx}$, using appropriate tensor powers of $\alth(\rx)$ and $\alth^{-1}(\rx)$.  $X$ is {\em bounded} if $\sup_{\rx\in \Sigma}\| X\|_{\rx}<\infty$.

If $U\subset \Sigma$ is open, we denote by ${\rm BT}^{p}_{q}(U, \altdelta)$ the Fr\'echet space of $(p,q)$ tensors on $U$, bounded with all covariant derivatives in the above sense. Among several equivalent definitions of manifolds of bounded geometry (see \cite{Sh2,bounded}), the one below is particularly useful in applications.

\begin{definition}\label{thp0.1}
A Riemannian manifold $(\Sigma,\alth)$ is of bounded geometry iff for each $\rx\in \Sigma$, there exists an open neighborhood of $\rx$, denoted $U_{\rx}$, and a smooth diffeomorphism
\[
\psi_{\rx}: U_{\rx} \xrightarrow{\sim} B_{d}(0,1)\subset \rr^{d}
\]
with $\psi_{\rx}(\rx)=0$, and such that if $\alth_{\rx}\defeq (\psi_{\rx}^{-1})^{*}\alth$ then:

\noindent
{\rm (C1)} the family $\{\alth_{\rx}\}_{\rx\in \Sigma}$ is  bounded in ${\rm BT}^{0}_{2}(B_{d}(0,1), \altdelta)$,

\noindent
 {\rm (C2)} there exists $c>0$ such that :
\[
c^{-1}\altdelta\leq \alth_{\rx}\leq c \altdelta, \ \rx\in \Sigma.
\]

 A family $\{U_{\rx}\}_{\rx\in \Sigma}$ resp. $\{\psi_{\rx}\}_{\rx\in \Sigma}$ as above will be called a family of {\em good chart neighborhoods}, resp. {\em good chart diffeomorphisms}.
\end{definition}

A known result (see \cite[Lemma 1.2]{Sh2}) says that one can find a covering $\Sigma=\bigcup_{i\in \nn}U_{i}$ by good chart neighborhoods $U_{i}= U_{\rx_{i}}$ $(\rx_{i}\in \Sigma)$ which is {\em uniformly finite}, i.e. there exists $N\in \nn$ such that $\bigcap_{i\in I}U_{i}= \emptyset$ if $\sharp I> N$. Setting $\psi_{i}= \psi_{\rx_{i}}$, we will call the sequence $\{U_{i}, \psi_{i}\}_{i\in \nn}$ a {\em good chart covering} of $\Sigma$. 

Furthermore, by \cite[Lemma 1.3]{Sh2} one can associate to a good chart covering a partition of unity:
\[
1=\sum_{i\in \nn}\chi_{i}^{2},  \ \ \chi_{i}\in \coinf(U_{i})
\]
such that $\{(\psi_{i}^{-1})^{*}\chi_{i}\}_{i\in \nn}$ is a bounded sequence in $\cinf_{\rm b}(B_{d}(0,1))$. Such a partition of unity will be called a {\em good partition of unity}.
\subsection{Bounded tensors and bounded diffeomorphisms}
\begin{definition}\label{defp0.2}
 Let $(\Sigma,\alth)$ be of  bounded geometry. We denote by ${\rm BT}^{p}_{q}(\Sigma,\alth)$ the spaces of  smooth $(q,p)$ tensors $X$ on $\Sigma$ such that if $X_{\rx}= (\exp_{\rx}^{\alth}\circ e_{\rx})^{*}X$, where $e_{\rx}: (\rr^{d}, \delta)\to (T_{\rx}\Sigma, h(\rx))$ is an isometry,  then the family 
 $\{X_{\rx}\}_{x\in \Sigma}$ is  bounded in ${\rm BT}^{p}_{q}(B_{d}(0, \frac{r}{2}), \altdelta)$. We equip ${\rm BT}^{p}_{q}(\Sigma, \alth)$ with its natural Fr\'echet space topology.
 
  We denote by $\cinfb(\rr; \BT^{p}_{q}(\Sigma, \alth))$ the space of smooth maps $\rr\in t\mapsto X(t)$ such that $\p_{t}^{n}X(t)$ is  uniformly bounded in $\BT^{p}_{q}(\Sigma, \alth)$ for $n\in \nn$. 
  
  We denote by $S^{\delta}(\rr; \BT^{p}_{q}(\Sigma, \alth))$, $\delta\in \rr$ the space of smooth maps $\rr\in t\mapsto X(t)$
  such that $\langle t\rangle^{-\delta+ n}\p_{t}^{n}X(t)$ is uniformly bounded in $\BT^{p}_{q}(\Sigma, \alth)$ for $n\in \nn$. 
 \end{definition}
It is well known (see e.g. \cite[Subsect. 2.3]{bounded}) that  we can replace in Def. \ref{defp0.2} the geodesic maps $\exp_{\rx}^{\alth}\circ e_{\rx}$ by $\psi_{\rx}^{-1}$, where $\{\psi_{\rx}\}_{\rx\in \Sigma}$ is any family of good chart diffeomorphisms as in Thm. \ref{thp0.1}. 
\begin{definition}\label{defdeboun}
 Let $(\Sigma, \alth)$ be an $n-$dimensional  Riemannian manifold of bounded geometry and $\chi: \Sigma\to \Sigma$ a smooth diffeomorphism. One says that $\chi$ is a {\em bounded diffeomorphism} of $(\Sigma, \alth)$ if for some some family of good chart diffeomorphisms $\{U_{\rx}, \psi_{\rx}\}_{\rx\in \Sigma}$,  
the maps
\[
\chi_{\rx}= \psi_{\chi(\rx)}\circ \chi\circ \psi_{\rx}^{-1}, \ \chi^{-1}_{\rx}= \psi_{\chi^{-1}(\rx)}\circ \chi^{-1}\circ \psi_{\rx}: B_{n}(0, 1)\to B_{n}(0,1)
\]
are bounded in $\cinfb(B_{n}(0,1))$ uniformly with respect to $\rx\in \Sigma$.
\end{definition}
It is easy to see that if the above properties are satisfied for some family of good chart  diffeomorphisms  then they are satisfied for any such family, furthermore bounded diffeomorphisms are stable under composition.

\subsection{Symbol classes}\label{symbolo}
We recall some well-known definitions about symbol classes on manifolds of bounded geometry, following \cite{Sh2, Ko, alnv1}.
\subsubsection{Symbol classes on $\rr^{n}$}\label{secp1.2.1}
Let $U\subset \rr^{d}$ be  an open set, equipped with the flat metric $\altdelta$ on $\rr^{d}$.

 We denote by $S^{m}(T^{*}U)$, $m\in \rr$,  the space of $a\in \cinf(U\times \rr^{d})$ such that
\[
 \langle \spexi\rangle^{-m+|\beta|}\p_{\rx}^{\alpha}\p_{k}^{\beta}a(\rx, \spexi)\hbox{  is  bounded on }U\times \rr^{d}, \ \forall \alpha, \beta\in \nn^{n}, 
\]
equipped with its canonical semi-norms $\| \cdot\|_{m,\alpha, \beta}$. 

We set 
\[
S^{-\infty}(T^{*}U)\defeq  \bigcap_{m\in \rr}S^{m}(T^{*}U),\ \ S^{\infty}(T^{*}U)\defeq  \bigcup_{m\in \rr}S^{m}(T^{*}U),
\]
 with their canonical Fr\'echet space topologies. If $m\in \rr$ and $a_{m-i}\in S^{m-i}(T^{*}U)$ we write
$a\simeq \sum_{i\in \nn}a_{m-i}$ if for each $p\in \nn$ 
\begin{equation}
\label{ep1.1}
r_{p}(a)\defeq  a- \sum_{i=0}^{p}a_{m-i}\in S^{m-p-1}(T^{*}U).
\end{equation}
It is well-known (see e.g. \cite[Sect. 3.3]{shubin}) that if $a_{m-i}\in S^{m-i}(T^{*}U)$, there exists $a\in S^{m}(T^{*}U)$, unique modulo $S^{-\infty}(T^{*}U)$ such that $a\simeq \sum_{i\in \nn}a_{m-i}$.

We denote by $S^{m}_{\rm h}(T^{*}U)\subset S^{m}(T^{*}U)$ the  space of $a$ such that $a(\rx, \lambda k)= \lambda^{m}a(\rx, \spexi)$, for $\rx\in U$, $|\spexi|\geq C$, $C>0$ and by $S^{m}_{\rm ph}(T^{*}U)\subset S^{m}(T^{*}U)$ the space of $a$ such that $a\simeq \sum_{i\in \nn}a_{m-i}$ for a sequence $a_{m-i}\in S^{m-i}_{\rm h}(T^{*}U)$ ($a$ is then called a \emph{poly-homogeneous}\footnote{These are also called classical symbols in the literature.} symbol). 
Following \cite{alnv1} one equips  $S^{m}_{\rm ph}(T^{*}U)$ with  the topology defined by  the semi-norms of $a_{m-i}$ in $S^{m-i}(T^{*}U)$ and $r_{p}(a)$ in $S^{m-p-1}(T^{*}U)$, (see \eqref{ep1.1}). This topology is strictly stronger than the topology induced by $S^{m}(T^{*}U)$.

The space  $S^{m}_{\rm ph}(T^{*}U)/S^{m-1}_{\rm ph}(T^{*}U)$ is isomorphic to $S^{m}_{\rm h}(T^{*}U)$, and the image of $a$ under the quotient map is called the {\em principal symbol} of $a$ and denoted by $\sigma_{\rm pr}(a)$. 

If $U= B_{n}(0,1)$ (more generally, if $U$ is relatively compact with smooth boundary), there exists a  continuous extension map $E: S^{m}(T^{*}U)\to S^{m}(T^{*}\rr^{d})$ such that $E a\traa{T^{*}U}=a$. Moreover $E$ maps $S^{m}_{\rm ph}(T^{*}U)$  into $S^{m}_{\rm ph}(T^{*}\rr^{d})$  and is continuous for the topologies of $S^{m}_{\rm ph}(T^{*}U)$ and $S^{m}_{\rm ph}(T^{*}\rr^{d})$, which means that all the maps
\[
a\mapsto (Ea)_{m-i},  \ \ a\mapsto r_{p}(Ea),
\]
are continuous.
\subsubsection{Time-dependent symbol classes on $\rr^{d}$}\label{timdep}
 We will also need to consider various classes of {\em time-dependent} symbols $a(t, \rx, \spexi)\in \cinf(\rr\times T^{*}U)$. 
 First of all the space $\cinf(\rr; S^{m}(T^{*}U))$  is defined as the 
  space of $a\in  \cinf(\rr\times T^{*}U)$ such that
 \[
 \langle \spexi\rangle^{-m+|\beta|}\p_{t}^{\gamma}\p_{\rx}^{\alpha}\p_{\spexi}^{\beta}a(t,\rx, \spexi)\hbox{  is bounded on }I\times U\times \rr^{d}, \ \forall \alpha, \beta\in \nn^{n}, \ \gamma\in \nn,
\]
for any interval $I\Subset \rr$.
We denote by  $\cinf_{\rm b}(\rr; S^{m}(T^{*}U))$  the subspace of symbols which are uniformly bounded   in $S^{m}(T^{*}U)$  with all time derivatives.

Furthermore, anticipating the need for some additional decay in $t$ in Sect. \ref{secscat}, we denote by $S^{\delta}(\rr; S^{m}(T^{*}U))$ the space of $a\in  \cinf(\rr\times T^{*}U)$ such that
 \[
\langle t\rangle^{-\delta+\gamma} \langle \spexi\rangle^{-m+|\beta|}\p_{t}^{\gamma}\p_{\rx}^{\alpha}\p_{\spexi}^{\beta}a(t,\rx, \spexi)\hbox{  is bounded on }\rr\times U\times \rr^{d}, \ \forall \alpha, \beta\in \nn^{n}, \ \gamma\in \nn.
\]
The  notation $a\sim \sum_i {a_{m-i}}$ and the poly-homogeneous spaces   
\[
\cinf_{({\rm b})}(\rr; S^{m}_{\rm ph}(T^{*}U)), \ \ S^{\delta}(\rr; S^{m}_{\rm ph}(T^{*}U)),
\]
are defined analogously, by requiring estimates on the time derivatives of the $a_{m-i}$ and $r_{p}$ in \eqref{ep1.1}.

\subsubsection{Symbol classes on $\Sigma$}\label{secp1.2.2}
Let $(\Sigma, \alth)$ be a Riemannian manifold of bounded geometry and $\{\psi_{\rx}\}_{\rx\in \Sigma}$ a family of good chart diffeomorphisms.
\begin{definition}\label{defp.1}
 We denote by $S^{m}(T^{*}\Sigma)$  for $m\in \rr$ the space of $a\in \cinf(T^{*}\Sigma)$ such that for each $\rx\in \Sigma$,  $a_{\rx}\defeq (\psi_{\rx}^{-1})^{*}a\in S^{m}(T^{*}B_{n}(0,1))$ and the family $\{a_{\rx}\}_{\rx\in \Sigma}$ is bounded
 in $S^{m}(T^{*}B_{n}(0,1))$.  We equip  $S^{m}(T^{*}\Sigma)$ with the semi-norms
 \[
\| a\|_{m, \alpha, \beta}= \sup_{\rx\in \Sigma}\| a_{\rx}\|_{m, \alpha, \beta}.
\]
Similarly we denote by $S^{m}_{\rm ph}(T^{*}\Sigma)$ the space of $a\in S^{m}(T^{*}\Sigma)$ such that for each $\rx\in \Sigma$, $a_{\rx}\in S^{m}_{\rm ph}(T^{*}B_{n}(0,1))$ and the family $\{a_{\rx}\}_{\rx\in \Sigma}$ is bounded
 in $S^{m}_{\rm ph}(T^{*}B_{n}(0,1))$. We equip $S^{m}_{\rm ph}(T^{*}\Sigma)$ with the semi-norms
 \[
\| a\|_{m, i, p, \alpha, \beta}= \sup_{\rx\in \Sigma}\| a_{\rx}\|_{m,i, p,\alpha, \beta}.
\]
where $\| \cdot \|_{m,i, p,\alpha, \beta}$ are the semi-norms defining the topology of $S^{m}_{\rm ph}(T^{*}B_{n}(0,1))$.

We also set $S^{\infty}_{({\rm ph})}(T^{*}\Sigma)= \bigcup_{m\in \rr}S^{m}_{({\rm ph})}(T^{*}\Sigma)$.
\end{definition}
The definition of $S^{m}(T^{*}\Sigma)$, $S^{m}_{\rm ph}(T^{*}\Sigma)$ and their  Fr\'echet space topologies are  independent on the choice of  the family $\{\psi_{\rx}\}_{\rx\in \Sigma}$ of good chart diffeomorphisms.

The notation $a\simeq \sum_{i\in \nn}a_{m-i}$ for $a_{m-i}\in S^{m-i}_{\rm ph}(T^{*}\Sigma)$ is defined as before.  If $a\in S^{m}_{\rm ph}(T^{*}\Sigma)$, we denote again by $a_{\rm pr}$ the image of $a$ in $S^{m}_{\rm ph}(T^{*}\Sigma)/S^{m-1}_{\rm ph}(T^{*}\Sigma)$.

The spaces $\cinf_{({\rm b})}(\rr; S^{m}_{({\rm ph})}(T^{*}\Sigma))$,  $S^{\delta}(\rr; S^{m}_{({\rm ph})}(T^{*}\Sigma))$ are defined as in \ref{timdep} and equipped with their  natural Fr\'echet space topologies.

\subsection{Sobolev spaces and smoothing operators}\label{sobolo}
Using the metric $\alth$ one defines the Sobolev spaces $H^{\sobo}(\Sigma)$ as follows.

\begin{definition} For $s\in \rr$ the {\em Sobolev space} $H^{\sobo}(\Sigma)$ is:
 \[
H^{\sobo}(\Sigma)\defeq \langle -\Delta_{\alth}\rangle^{-\sobo/2}L^{2}(\Sigma),
\]
with its natural Hilbert space topology, where $-\Delta_{\alth}$ is the Laplace-Beltrami operator on $(\Sigma,\alth)$, strictly speaking the closure of its restriction to $\coinf(\Sigma)$. 
\end{definition}

We further set 
\[
H^{\infty}(\Sigma)\defeq \textstyle\bigcap_{m\in \zz}H^{m}(\Sigma), \quad H^{-\infty}(\Sigma)\defeq \textstyle\bigcup_{m\in \zz}H^{m}(\Sigma),
\]
equipped with their Fr\'echet space topologies.  

We denote by $\cW^{-\infty}(\Sigma)$ the Fr\'echet space  $B(H^{-\infty}(\Sigma), H^{\infty}(\Sigma))$ with its Fr\'echet space topology, given by the semi-norms 
\[
\| a\|_{m}= \| a\|_{B(H^{-m}(\Sigma),  H^{m}(\Sigma))}, \ \ m\in \nn.
\]
This allows us to define $\cinf_{({\rm b})}(\rr; \cW^{-\infty}(\Sigma))$,  $S^{\delta}(\rr; \cW^{-\infty}(\Sigma))$, the latter consisting of operator-valued functions $a(t)$ such that
\[
\|\p_{t}^{\gamma}a(t)\|_{m}\in O(\langle t\rangle^{\delta- \gamma}), \ \ \forall \gamma, m\in \nn.
\]
\subsection{Pseudodifferential operators}\label{pdosec}
Starting from the well-known Weyl quantization on open subsets of $\rr^d$, one constructs a quantization map $\Op$ for symbols in $S^m(T^*\Sigma)$ using a  good chart covering of $\Sigma$ and good chart diffeomorphisms. More precisely let $\{U_{i}, \psi_{i}\}_{i\in \nn}$ be a good chart covering of $M$   and  
\[
\sum_{i\in \nn}\chi_{i}^{2}= \one
\]
a subordinate good partition of unity, see 
 Subsect. \ref{ss:mbg}. If
\[
(\psi_{i}^{-1})^{*}dg\eqdef  m_{i}dx,
\]
we set
\[
\begin{array}{rl}
T_{i}:&L^{2}(U_{i}, dg)\to L^{2}(B_{n}(0,1), dx),\\[2mm]
&u\mapsto m_{i}^{\12}(\psi_{i}^{-1})^{*}u,
\end{array}
\]
so that $T_{i}:L^{2}(U_{i}, dg)\to L^{2}(B_{n}(0,1), dx)$ is unitary.  We then fix an extension map 
\[
E: S^{m}_{\rm ph}(T^{*}B_{d}(0, 1))\to S^{m}_{\rm ph}(T^{*}\rr^{d}).
\]
\begin{definition}\label{def-de-op}
 Let $a= a(t)\in \cinf(\rr; S^{m}_{\rm ph}(T^{*}M))$. We set
 \[
\Op(a)\defeq  \sum_{i\in \nn}\chi_{i}T_{i}^{*}\circ \Op^{\rm w}(Ea_{i})\circ T_{i}\chi_{i},
\]
where $a_{i}\in S^{m}_{\rm ph}(T^{*}B_{d}(0,1))$ is the push-forward of $a\traa{T^{*}U_{i}}$ by $\psi_{i}$ and $\Op^{\rm w}$ is the Weyl quantization.
\end{definition}

 If $\Op'$ is another such quantization map for different choices of $U_{i}, \psi_{i}, \chi_{i}$ and $E$  then
\begin{flalign*}
&S^{m}_{\rm ph}(T^{*}\Sigma)\to \cW^{-\infty}(\Sigma)\\[2mm]
\Op - \Op': \ \ & \cinf_{({\rm b})}(\rr; S^{m}_{\rm ph}(T^{*}\Sigma))\to \cinf_{({\rm b})}(\rr; \cW^{-\infty}(\Sigma)),\\[2mm]
&S^{\delta}(\rr; S^{m}_{\rm ph}(T^{*}\Sigma))\to S^{\delta}(\rr; \cW^{-\infty}(\Sigma)),
\end{flalign*}
are bounded. Then  one defines the classes
 \[
\bea 
\Psi^{m}(\Sigma)&\defeq \Op (S^{m}_{\rm ph}(T^{*}\Sigma))+ \cW^{-\infty}(\Sigma),\\[2mm]
\cinf_{({\rm b})}(\rr; \Psi^{m}(\Sigma))&\defeq \Op(\cinf_{({\rm b})}(\rr; S^{m}_{\rm ph}(T^{*}\Sigma)))+ \cinf_{({\rm b})}(\rr; \cW^{-\infty}(\Sigma)),\\[2mm]
S^{\delta}(\rr; \Psi^{m}(\Sigma))&\defeq \Op (S^{\delta}(\rr; \Psi^{m}_{\rm ph}(T^{*}\Sigma))+ S^{\delta}(\rr; \cW^{-\infty}(\Sigma)).
\eea 
\]
Thanks to including the  ideal   $\cW^{-\infty}(\Sigma)$ of smoothing operators, the so-obtained pseudodifferential classes are stable under composition,  for example  
\[
\Psi^{m_1}(\Sigma)\circ \Psi^{m_2}(\Sigma)\subset \Psi^{m_1+m_2}(\Sigma).
\]

Note that $S^{\delta}(\rr;\Psi^{m}(\Sigma))= \langle t\rangle^{\delta}S^{0}(\rr; \Psi^{m}(\Sigma))$ and similarly with $\Psi^{m}(\Sigma)$ replaced by $\cW^{-\infty}(\Sigma)$  so in what follows one can assume without loss of generality that  $\delta=0$.

The spaces $\cW^{-\infty}(\Sigma)$, $\cinf_{({\rm b})}(\rr; \cW^{-\infty}(\Sigma))$ and $S^{\delta}(\rr; \cW^{-\infty}(\Sigma))$ have natural Fr\'e\-chet space topologies. If necessary we equip the spaces $\Psi^{m}(\Sigma)$, $\cinf_{({\rm b})}(\rr; \Psi^{m}(\Sigma))$ and $S^{\delta}(\rr; \Psi^{m}(\Sigma))$ with the quotient topology obtained from the map:
\[
(c, R)\mapsto \Op(c)+R
\]
between the appropriate spaces.

If $a\in \Psi^{m}(\Sigma)$, the {\em principal symbol} $\sigma_{\rm pr}(a)\in S^{m}_{\rm h}(T^{*}\Sigma)$ is defined in analogy to the case $\Sigma=\rr^d$. The operator $a$ is {\em elliptic} if  there exists $C>0$ such that
\beq\label{eq:elliptic}
| \sigma_{\rm pr}(a)|\geq C |\spexi|^{m}, \ \ |\spexi|\geq 1,
\eeq
uniformly in the chart open sets. If $a\in \cinf(\rr; \Psi^{m}(\Sigma))$ we say that $a$ is {\em  elliptic} if $a(t)$ is elliptic for all $t\in\rr$ and the constant $C$ in (\ref{eq:elliptic}) is locally uniform in $t$. For $a\in \cinf_{\rm b}(\rr; \Psi^{m}(\Sigma))$ or $S^{0}(\rr; \Psi^{m}(\Sigma))$ there is also a corresponding notion of ellipticity, where we require $C$ to be uniform in $t$.

As shown in \cite{bounded}, the pseudodifferential classes $\Psi^m(\Sigma)$ fit into the general framework of Ammann, Lauter, Nistor and Vasy \cite{alnv1}, and consequently they have many convenient properties that generalize well-known facts for say, pseudodifferential operators on closed manifolds, such as the existence of complex powers for elliptic, bounded from below operators. 

We state below a particular case of {\em Seeley's theorem} for real powers, partly proved in \cite[Sect. 5]{bounded} and based on a general result from \cite{alnv1}.
\begin{theorem}[Seeley's theorem]\label{seeley}
Let $a\in \cinf(\rr; \Psi^{m}(\Sigma))$ be elliptic, selfadjoint with $a(t)\geq c(t)\one$, $c(t)>0$. Then $a^{\alpha}\in \cinf(\rr; \Psi^{m\alpha}(\Sigma))$ for any $\alpha\in \rr$ and $\sigma_{\rm pr}(a^{\alpha})(t)= \sigma_{\rm pr}(a(t))^{\alpha}$.

The same result holds replacing  $\cinf(\rr; \Psi^{m}(\Sigma))$ by $\cinf_{({\rm b})}(\rr; \Psi^{m}(\Sigma))$ or also by $S^{0}(\rr; \Psi^{m}(\Sigma))$ if one assumes $a(t)\geq c_{0}\one$ for $c_{0}>0$.
\end{theorem}
\proof The $\cinf_{({\rm b})}$ cases  are proved in \cite[Thm. 5.12]{bounded}, by checking that the general framework of \cite{alnv1} applies to these two situations.  The $S^{0}$ case can be proved similarly. The only point deserving special care is the {\em spectral invariance} of the ideal $S^{0}(\rr; \cW^{-\infty}(\Sigma))$, which we explain in some detail. Let $r_{-\infty}\in S^{0}(\rr; \cW^{-\infty}(\Sigma))
$, considered as a bounded operator  on $L^{2}(\rr_{t}\times \Sigma_{\rx})$. The spectral invariance property is the fact that if $\one - r_{-\infty}$ is invertible in $B(L^{2}(\rr_{t}\times \Sigma_{\rx}))$ then $(\one - r_{-\infty})^{-1}= \one- r_{1, -\infty}$ for $r_{1, -\infty}\in S^{0}(\rr; \cW^{-\infty}(\Sigma))$.
This can be however proved exactly as in \cite[Lemma 5.5]{bounded}. \qed

\subsection{Egorov's theorem} If $b(t)\in C^\infty(\rr;\Psi^m(\Sigma))$ (or more generally, if $b(t)$ is a square matrix consisting of elements of $C^\infty(\rr;\Psi^m(\Sigma))$ and $H^{-\infty}(\Sigma)$ is tensorized by powers of $\cc$ accordingly) we denote by 
\[
\cU_{b}(t,s): H^{-\infty}(\Sigma)\to H^{-\infty}(\Sigma)
\]
the evolution generated by $b(t)$, i.e. the Cauchy evolution of $\pe_t-\i b(t)$, or put in other words, the unique solution (if it exists) of the system
\beq\label{eq:defUb}
\begin{cases}
\frac{\p}{\p t}\cU_{b}(t,s)= \i  b(t)\cU_{b}(t,s),\\[2mm]
\frac{\p}{\p s}\cU_{b}(t,s)= -\i  \cU_{b}(t,s)b(s),\\[2mm]
\cU_{b}(t,s)=\one.
\end{cases}
\eeq
The existence of $\cU_{b}(t,s)$ can typically be established if $b(t)$ defines a differentiable family of self-adjoint operators on a Hilbert space, or a small perturbation of such family. Specifically, consider $b(t)\in \cinf(\rr; \Psi^{1}(\Sigma))$ such that $b(t)=b_1(t)+b_0(t)$ with $b_i(t)\in \cinf(\rr; \Psi^{i}(\Sigma))$ and:
\[
({\rm E})\quad b_1(t)\hbox{ is  elliptic and bounded from below on } H^\infty(\Sigma), \hbox{ locally uniformly in }t.
\]
Using \cite[Prop. 2.2]{alnv1} it follows that $b(t)$ is closed with domain $\Dom b(t)= H^{1}(\Sigma)$. Moreover  the map $\rr\ni t\mapsto b(t)\in B(H^{1}(\Sigma), L^{2}(\Sigma))$ is norm continuous. It follows that we can define $\cU_{b}(t,s)$, using for instance \cite[Thm. X.70]{RS}. In the present setup one can prove a result known generally as Egorov's theorem, we refer to \cite{bounded} for the details and proofs.

\begin{lemma}\label{ego.1}
Assume $({\rm E})$. Then:
\ben
\item 
 $\cU_{b}(t,s)\in B(H^{m}(\Sigma))$ for $m\in \rr$ or $m= \pm \infty$.  
 \item  if $r\in \cW^{-\infty}(\Sigma)$ then
$\cU_{b}(t,s)r, \ r\cU_{b}(s,t)\in \cinf(\rr^{2}_{t,s}, \cW^{-\infty}(\Sigma))$.

 \item if moreover $b(t)\in S^{0}(\rr; \Psi^{1}(\Sigma))$ and $b(t)- b^{*}(t)\in S^{-1- \delta}(\rr; \Psi^{0}(\Sigma))$ for $\delta>0$ then 
 $\cU_{b}(t,s)$ is uniformly bounded in $B(L^{2}(\Sigma))$.
 \een
\end{lemma}
\begin{theorem}[Egorov's theorem] 
Let $c\in \Psi^{m}(\Sigma)$ and $b(t)$ satisfying $({\rm E})$. Then
\[
c(t,s)\defeq  \cU_{b}(t,s)c \cU_{b}(s,t)\in \cinf(\rr^{2}_{t,s}, \Psi^{m}(\Sigma)).
\]
Moreover 
\[
\sigma_{\rm pr}(c)(t,s)= \sigma_{\rm pr}(c)\circ \Phi(s,t),
\]
where $\Phi(t,s): T^{*}\Sigma\to T^{*}\Sigma$ is the flow of the time-dependent Hamiltonian $\sigma_{\rm pr}(b)(t)$.
\end{theorem}

\subsection{Some auxiliary results} \label{ss:auxr}
For the sake of having a slightly more short-hand notation, for  $(\Sigma, \alth)$ of bounded geometry we set: 
\[
\bea
\Psi_{\td}^{m, \delta}(\rr; \Sigma) &\defeq S^{\delta}(\rr; \Psi^{m}(\Sigma)).
\eea
\]
for pseudodifferential operator classes with time decay $(\td)$ of the symbols. 

\subsubsection{Difference of fractional powers}
We now state an auxiliary result about fractional powers of elliptic operators that will be needed later on.
\begin{proposition}\label{l5.1}
 Let   $a_{i}\in\Psi_{\td}^{2,0}(\rr; \Sigma)$, $i=1,2$ elliptic with $a_{i}=a_{i}^{*}$ and $a_{i}(t)\geq c_{0}\one$ for some $c_{0}>0$. Assume that $a_{1}- a_{2}\in \Psi_{\td}^{2, -\delta}(\rr; \Sigma)$ with $\delta>0$. Then  for each $\alpha\in \rr$ one has:
 \[
 a_{1}^{\alpha}- a_{2}^{\alpha}\in \Psi_{\td}^{2\alpha, -\delta}(\rr;\Sigma).  
 \]
 \end{proposition}

Prop. \ref{l5.1} is proved in Subsect. \ref{ssecap1}.
\subsubsection{Ressummation of symbols}
We now examine the ressummation of symbols.  One can think of this as a statement about the uniform symbol classes on $\rr^{d}$, after applying a chart diffeomorphism. 

We denote $\Psi^{-\infty,-\delta}_{\td}(\rr; \Sigma)\defeq\bigcap_{m\in\rr}\Psi^{m,-\delta}_{\td}(\rr; \Sigma)$.

\begin{lemma}\label{l5.2}
Let $\delta\in \rr$ and let $(m_{j})$ be a real sequence decreasing to $-\infty$. Then
if  $a_{j}\in \Psi_{\td}^{m_{j},-\delta}(\rr; \Sigma)$ there exists $a\in \Psi_{\td}^{ m_{0},-\delta}(\rr; \Sigma)$, unique modulo $\Psi^{-\infty,-\delta}_{\td}(\rr; \Sigma)$, such that
 \[
 a\sim \sum_{j=0}^{\infty}a_{j}, \hbox{ i.e. }
 a-\sum_{j=0}^{N}a_{j}\in\Psi_{\td}^{ m_{N+1},-\delta}(\rr; \Sigma), \ \forall N\in \nn.
  \]
 \end{lemma}
\proof By introducing the new variable $s=\int_{0}^{t}\langle \sigma\rangle^{-1}d\sigma$ (so that $\langle t\rangle \p_{t}= \p_{s}$) and putting the extra variable $s$ together with the $\bx$ variables we can reduce ourselves to the situation covered by the standard proof (see e.g. \cite[Prop. 3.5]{shubin}).\qed

\section{Parametrix for the Cauchy evolution and Hadamard states}\init\label{sec2}

\subsection{Model Klein-Gordon equation}\label{sec2.1}

We fix a $d-$dimensional manifold $\Sigma$ equipped with a reference Riemannian metric $\altk$  such that $(\Sigma, \altk)$ is of bounded geometry. We equip $M= \rr\times \Sigma$, the elements of which are denoted by $x= (t, \rx)$, with a  Lorentzian metric  $\altg$ and a real function $\altV$ such that:
\beq\label{as:FLRW}
\begin{array}{l}
\altg= - dt^{2}+ \alth_{ij}(t,\rx )d\rx^{i}d\rx^{j},\\[2mm]
\alth\in \cinf(\rr, \BT^{0}_{2}(\Sigma, \altk)), \ \alth^{-1}\in \cinf(\rr; \BT^{2}_{0}(\Sigma, \altk)),\\[2mm]
\altV\in \cinf(\rr; \BT^{0}_{0}(\Sigma, \altk)).
\end{array}
\eeq
Although the first assumption may look restrictive, we will give in Subsect. \ref{s10.1} a reduction procedure that will allow us to treat more general cases.

In this setup, the Klein-Gordon operator $P=-\Box_\altg+\altV$ equals
\beq\label{eq:modelP}
\bea
P &= |\alth|^{-\12}\pe_{t}|\alth|^{\12}\pe_{t}- |\alth|^{-\12}\pe_{i}\alth^{ij}|\alth|^{\12}\pe_{j}+\altV \\
   &= \pe_{t}^{2}+ r(t,\rx )\pe_{t}+ a(t, \rx, \pe_{\rx}),
\eea
\eeq
where
\[
a(t, \rx, \pe_{\rx})= - |\alth|^{-\12}\pe_{i}\alth^{ij}|\alth|^{\12}\pe_{j}+ \altV(t,\rx ) 
\]
is formally self-adjoint with respect to the $t$-dependent $L^{2}(\Sigma, |\alth|^{\12}dx)$-inner product and
\[
r(t,\rx )= |\alth|^{-\12}\p_{t}(|\alth|^{\12})(t,\rx ).
\]
Note that the above function is closely related to the extrinsic curvature of $\Sigma$ in $M$. 

In the sequel we will often abbreviate $a(t, \rx, \pe_{\rx})$ by $a(t)$ or $a$.

\subsection{Construction of parametrix}\label{sec2.3}

Following \cite{bounded} we now explain how one obtains a parametrix for the Cauchy evolution for the model Klein-Gordon operator (and a splitting of it) by means of an approximate time-dependent diagonalization.  We will then adapt it to the setup of scattering theory. 

The first step consists of observing that the Klein-Gordon equation $(\p_t^2+ r(t)\p_t + a(t))\phi(t)=0$ is equivalent to
\beq\label{eq:evA}
\i^{-1}\p_t \psi(t) = \AH(t) \psi(t), \quad \mbox{ where \ } \AH(t)=\mat{0}{\one}{a(t)}{\i r(t)}, 
\eeq
by setting
\beq\label{eq:idpsi}
\psi(t)=\begin{pmatrix}\phi(t) \\ \i^{-1}\p_t \phi(t)\end{pmatrix}.
\eeq
Let us  denote by $\cU(s,t)$ the evolution generated by $\AH(t)$, cf. (\ref{eq:defUb}). 
Recall that on Cauchy data on $\Sigma_{s}= \{s\}\times \Sigma$, we have a symplectic form induced from an operator $G(s)$,  defined by:
\[
G= (\varrho_{s}G)^{*}\circ G(s)\circ (\varrho_{s}G).
\]
Here the formal adjoint will always be taken with respect to the $L^{2}(\Sigma, |\alth|^{\12}dx)$-inner product. We have also introduced the hermitian operator $q(s)= \i G(s)$. It is well known that with these choices, $q(s)$ equals specifically
\beq\label{defdeq(s)}
q(s)= \mat{0}{\one}{\one}{0},
\eeq
in particular it does not depend on $s$ (we will thus simply write $q$ instead). Furthermore,
\beq\label{eq-e2}
\cU^{*}(t,s)q\cU(t,s)= q,
\eeq
(the Cauchy evolution is symplectic).
\subsubsection{Riccati equation}\label{sec2.2} The approximate diagonalization of $\cU(s,t)$ will be based on solving the Riccati equation
\begin{equation}
\label{enp.11c}
\i \p_{t}b- b^{2}+a + \i rb=0,
\end{equation}
modulo smoothing terms, where the unknown is $b(t)\in C^{\infty}(\rr;\Psi^1(\Sigma))$. By repeating the arguments in \cite{GW,GW2} this can be solved modulo terms in $\cinf(\rr; \cW^{-\infty}(\Sigma))$. Concretely, supposing for the moment that $a(t)\geq c(t)\one$  for $c(t)>0$, upon setting $\epsilon= a^{\12}$, $b= \epsilon+ b_{0}$ one obtains the equations:
\[
\bea 
b_{0} &= \frac{\i }{2}(\epsilon^{-1}\p_{t}\epsilon+ \epsilon^{-1}r \epsilon)+ F(b_{0}),\\[2mm]
F(b_{0}) &= \12 \epsilon^{-1}(\i \p_{t}b_{0}+ [\epsilon, b_{0}]+ \i r b_{0}- b_{0}^{2}).
\eea 
\]
These can be solved  by substituting a poly-homogeneous expansion of the symbol of $b_0$, yielding an approximate solution of (\ref{enp.11c}) in the sense that
\beq\label{eq:ricattiwrem}
\i \p_{t}b- b^{2}+a + \i rb=r_{-\infty}\in \cinf(\rr; \cW^{-\infty}(\Sigma)).
\eeq
 Set 
\beq
b^{+}=b, \quad b^{-}= - b^{*}.
\eeq
Taking the adjoint of both sides of (\ref{eq:ricattiwrem}) with respect to the $t$-dependent inner product $L^{2}(\Sigma, |\alth|^{\12}dx)$ and using that
 \[
( \p_{t}b)^{*}= \p_{t}(b^{*})+ r b^{*}- b^{*}r,
 \]
 we obtain
 \beq\label{eq-ric}
 \i \p_{t}b^{\pm}- b^{\pm 2}+ a + \i rb^{\pm}=r_{-\infty}^{\pm},
 \eeq
 with $r^{+}_{-\infty}= r_{-\infty}$, $r^{-}_{-\infty}= r_{-\infty}^{*}\in\cinf(\rr; \cW^{-\infty}(\Sigma))$.  
 
In general  we can find a cutoff function $\varphi\in \coinf(\rr)$ such that $a(t)+ \varphi(a(t))\geq c(t)\one$ for $c(t)$ as above,  using the locally uniform ellipticity of $a(t)$. Since $\varphi(a(t))$ is a smoothing operator,  replacing $a(t)$ by $a(t)+ \varphi(a(t))$ is a harmless modification.

 A redefinition of $b(t)$ involving a cutoff in low frequencies as in \cite{GW2,bounded} gives then control of the norm sufficient to obtain in addition
\beq\label{eq:addi}
(b^+(t)- b^{-}(t))^{-1}\geq C(t) \epsilon(t)^{-1}
\eeq
for some $C(t)>0$, while keeping the property that $b^\pm(t)=\pm\epsilon(t)+\cf(\rr;\Psi^0(\Sigma))$, and with  \eqref{eq-ric} still valid for some $r_{-\infty}^{\pm}\in \cinf(\rr; \cW^{-\infty}(\Sigma))$.

Observe now that the Riccati equation \eqref{eq-ric} implies the following approximate factorization of the Klein-Gordon operator:
\begin{equation}
\label{eq-e1}
(\pe_{t}+ \i b^{\pm}(t)+ r(t))\circ (\pe_{t}- \i b^{\pm}(t))= \pe_{t}^{2}+ r\pe_t + a   - r_{-\infty}^{\pm}.
\end{equation}
Such a factorization of the Klein-Gordon operator was already used  by Junker \cite{junker} and Junker and Schrohe \cite{junker} to construct pure  Hadamard states in the case the Cauchy surface $\Sigma$ is compact. 

 Here we use (\ref{eq-e1}) to diagonalize (\ref{eq:evA}) by setting
\[
\tilde\psi(t)\defeq \begin{pmatrix}\pe_t-\i b^{-}(t) \\ \pe_t-\i b^{+}(t)\end{pmatrix}\phi(t).
\]
A direct computation yields then $\tilde{\psi}(t)= S^{-1}(t)\psi(t)$ with
 \beq\label{e4.01}
 S^{-1}(t)= \i \mat{- b^{-}(t)}{\one}{-b^{+}(t)}{\one}, \ \  S(t)= \i^{-1}\mat{\one}{-\one}{b^{+}(t)}{-b^{-}(t)}(b^{+}(t)- b^{-}(t))^{-1},
 \eeq
where well-definiteness and invertibility of $S(t)$ rely on the fact that $b^{+}(t)- b^{-}(t)$ is invertible by \eqref{eq:addi}. We obtain from \eqref{eq-e1} that
  \[
  \bea
& \mat{\pe_{t}+ \i b^{-}+ r}{0}{0}{\pe_{t}+ \i b^{+}+r}\tilde{\psi}(t)
 =\begin{pmatrix}
 \pe_{t}^{2}+ a+r\pe_{t}- r_{-\infty}^{-}\\ \pe_{t}^{2}+ a+ r\pe_{t}- r_{-\infty}^{+}\end{pmatrix}\phi(t)\\[2mm]
 &=\mat{r_{-\infty}^{-}}{0}{r_{-\infty}^{+}}{0}S(t)\tilde{\psi}(t)
 = \i^{-1}\mat{r_{-\infty}^{-}}{-r_{-\infty}^{-}}{r_{-\infty}^{+}}{-r_{-\infty}^{+}}(b^{+}- b^{-})^{-1}\tilde{\psi}(t).
  \eea
 \]
Therefore, $\tilde\psi(t)$ solves a diagonal matrix equation modulo smooth terms. More precisely, we have $\tilde{\psi}(t)= \cU_{B}(t, s)\tilde{\psi}(s)$ for
\beq\label{e4.02}
B(t)=\tilde{B}(t)+ R_{-\infty}(t),
\eeq
\beq\label{e4.02p}
\tilde{B}(t)= \mat{-b^{-}+ \i r}{0}{0}{-b^{+}+\i r}, \quad R_{-\infty}(t)=- \mat{r_{-\infty}^{-}}{-r_{-\infty}^{-}}{r_{-\infty}^{+}}{-r_{-\infty}^{+}}(b^{+}- b^{-})^{-1},
\eeq
Ultimately, we can thus conclude that
\beq\label{e4.00}
\cU(t,s)= S(t)\cU_{B}(t,s)S(s)^{-1}.
\eeq
\subsection{Improved approximate diagonalization}\label{ss:iad}
It is convenient to modify $S(t)$ to obtain a simple formula for the symplectic form $S^{*}(t)q(t)S(t)$ preserved by the almost diagonalized evolution.  Namely, setting
\beq\label{eq:defT}
\bea
&T(t)\defeq S(t)(b^{+}- b^{-})^{\12}(t)= \i^{-1}\mat{\one}{-\one}{b^{+}}{-b^{-}}(b^{+}- b^{-})^{-\12}, \\
&T^{-1}(t)= \i (b^{+}- b^{-})^{-\12}\mat{-b^{-}}{\one}{-b^{+}}{\one},
\eea
\eeq
we find that for $q$ defined in \eqref{defdeq(s)} one has:
\beq\label{hamon}
T^{*}(t) q T(t)=\mat{\one}{0}{0}{-\one}\eqdef q^{\adg}.
\eeq
We now define 
\beq\label{intolo}
\cU(t,s)\eqdef T(t)\cU^{\adg}(t,s)T(s)^{-1}, 
\eeq
and we obtain that $\cU^{\adg}(t,s)^{*}q^{\adg} \cU^{\adg}(t,s)=q^{\adg}$. Furthermore, the generator of the evolution group $\{\cU^{\adg}(t,s)\}_{t,s\in\rr}$ is:
\beq\label{eq:defH}
\bea
H^{\adg}(t)&=(b^{+}- b^{-})^{-\12}B(t)(b^{+}- b^{-})^{\12}-\i \p_{t}(b^{+}- b^{-})^{-\12}(b^{+}- b^{-})^{\12}\\
&=\mat{- b^{-}+ r_{b}^-}{0}{0}{-b^{+}+ r_{b}^+}- (b^{+}- b^{-})^{-\12}\mat{r_{-\infty}^{-}}{-r_{-\infty}^{-}}{r_{-\infty}^{+}}{-r_{-\infty}^{+}}(b^{+}- b^{-})^{-\12},
\eea
\eeq
where $r^\pm_{-\infty}\in C^\infty(\rr;\cW^{-\infty}(\Sigma))$ are the remainder terms from \eqref{eq-ric}, and
\beq\label{eq:rbpm}
r_{b}^\pm= \i r+ [(b^{+}- b^{-})^{-\12}, b^{\pm}]-\i \p_{t}(b^{+}- b^{-})^{-\12}(b^{+}- b^{-})^{\12}\in \Psi^{0}(\Sigma).
\eeq  
This way, denoting by $H^{\dg}$ the diagonal part of $H^{\adg}(t)$, using that $H^{\adg}(t)^{*}q^{\adg}= q^{\adg} H^{\adg}(t)$ we have:
\[
H^{\dg}(t)= H^{\dg*}(t), \ \ H^{\dg}(t)= \mat{\epsilon^{+}(t)}{0}{0}{\epsilon^{-}(t)},
\]
where
\[
\epsilon^{\pm}= - b^{\mp}+ r_{b}^{\mp} +\cinf(\rr; \cW^{-\infty}(\Sigma)), 
\]
and 
$H^{\adg}(t)=H^{\dg}(t)+ V^{\adg}_{-\infty}(t)$, where $V^{\adg}_{-\infty}(t)\in \cinf(\rr; \cW^{-\infty}(\Sigma)\otimes B(\cc^2))$. The evolution $\cU^{\dg}(t,s)$ generated by $H^{\dg}(t)$ is diagonal, in fact:
\beq\label{e4.103}
\cU^{\dg}(t,s)= \mat{\cU_{\epsilon^+}(t,s)}{0}{0}{\cU_{\epsilon^-}(t,s)}.
\eeq
Moreover:
\begin{equation}
\label{turlututu}
\bea
\cU(t,s)&=T(t)\cU^{\adg}(t,s)T(s)^{-1}\\
&= T(t)\cU^{\dg}(t,s)T(s)^{-1}+ \cinf(\rr^{2}; \cW^{-\infty}(\Sigma)).
\eea
\end{equation}
This is shown   by an `interaction picture' argument explained in detail in \cite{bounded}; we omit the proof here.

\begin{remark}\label{remrem} One easily sees  that $S(t)$ is an isomorphism from  $L^{2}(\Sigma)\oplus L^{2}(\Sigma)$ to $H^{1}(\Sigma)\oplus L^{2}(\Sigma)$ (the so-called {\em energy space}  of Cauchy data of \eqref{eq:evA}), while $T(t)$ is an isomorphism from $L^{2}(\Sigma)\oplus L^{2}(\Sigma)$  to 
 $H^{\12}(\Sigma)\oplus H^{-\12}(\Sigma)$ (this is the {\em charge space} that appears naturally in the quantization of the Klein-Gordon equation).
 \end{remark}

\subsection{Splitting of the parametrix and of the Cauchy evolution}\label{sec2.5}
Let us set
\beq\label{defdepiplusmoins}
\pi^{+}= \mat{\one}{0}{0}{0}, \ \ \pi^{-}=\mat{0}{0}{0}{\one}.
\eeq
Since $\cU^{\dg}(t,s)$ is diagonal  we have:
\[
\cU^{\dg}(t,s)= \cU^{\dg}(t,s) \pi^++ \cU^{\dg}(t,s)\pi^{-},
\] 
with $\cU^{\dg}(t,s)\pi^\pm$ propagating with wave front set contained in $\cN^\pm$ (this follows from $b^\pm$ being $\pm\epsilon$ modulo terms of lower order). 
This suggests that at least modulo smoothing terms, the splitting of $\cU(t,s)$ at time $s$ should be given by a pair of operators $c_{\rf}^\pm(s)$ defined as follows. We first fix a reference time $t_0\in\rr$.

\begin{definition}\label{defdec} We set:
 \[
c_{\rf}^{\pm}(t_0)\defeq  T(t_0)\pi^{\pm}T^{-1}(t_0)=  \mat{\mp(b^{+}- b^{-})^{-1}b^{\mp}}{\pm(b^{+}- b^{-})^{-1}}{\mp b^{+}(b^{+}- b^{-})^{-1}b^{-}}{\pm b^{\pm}(b^{+}- b^{-})^{-1}}(t_0).
\]
\end{definition}
Then $c_{\rf}^{\pm}(t_0)$ is a $2\times 2$ matrix of pseudo\-differential operators and 
\[
c_{\rf}^{\pm}(t_0)^{2}= c_{\rf}^{\pm}(t_0), \  \ c_{\rf}^{+}(t_0)+ c_{\rf}^{-}(t_0)= \one.
\]
We set:
\beq\label{e2.10}
\cU^{\pm}(t,s)\defeq  \cU(t,t_0 )c_{\rf}^{\pm}(t_0)\cU(t_0,s),
\eeq
so that 
\beq\label{e2.11}
\cU(t,s)= \cU^{+}(t,s)+ \cU^{-}(t,s).
\eeq
This splitting has the following properties (see \cite{bounded}):

\begin{proposition}\label{newprop.2}
 \beq\label{eq-e-2}
\bea 
i)& \ \cU^{\pm}(t, s)\cU^{\pm}(s,t')= \cU^{\pm}(t, t'),\\[2mm]
ii)&\ (\pe_{t}- \i \AH(t))\cU^{\pm}(t,s)= \cU^{\pm}(t,s)(\pe_{s}+ \i \AH(s))=0,\\[2mm]
iii)& \ \WF(\cU^{\pm}(t,s))'=\{(X, X')\in T^{*}\Sigma\times T^{*}\Sigma : \  X= \Phi^{\pm}(t,s)(X')\},
\eea 
\eeq
where $\Phi^{\pm}(t,s): T^{*}\Sigma\to T^{*}\Sigma$ is the symplectic flow generated by the time-dependent Hamiltonian $\pm(\alth^{ij}(t,\rx )\spexi_{i}\spexi_{j})^{\12}$.
\end{proposition}

If we set for $t\in\rr$:
\beq\label{eq-e3}
\cU^{\pm}(t,t)\eqdef  c^{\pm}_{\rf}(t)=\cU(t, t_0)c_{\rf}^{\pm}(t_0)\cU(t_0,t),
\eeq
then 
\[
c^{\pm}_{\rf}(t)^{2}= c^{\pm}_{\rf}(t), \quad c^{+}_{\rf}(t)+ c^{-}_{\rf}(t)= \one, \quad c^{\pm}_{\rf}(t)= \cU(t,s)c^{\pm}_{\rf}(s)\cU(s,t).
\]
Moreover from \eqref{hamon} and the fact that $\pm q^{\rm ad}\circ \pi^{\pm}\geq 0$, we obtain that 
\[
\lambda^{\pm}_{\rf}(t)=\pm q\circ c^{\pm}_{\rf}(t)\geq 0.
\]
As a consequence $c^\pm_\rf(t)$ are  the time-$t$ covariances of a Hadamard state \cite{bounded}. In general, we say that a state is a \emph{regular Hadamard state} if its time-$t$ covariances differ from $c_{\rf}^\pm(t)$ by terms in $\cW^{-\infty}(\Sigma)\otimes B(\cc^2)$, and one can show that it suffices to check that property for one value of $t$ \cite{bounded}. In summary:

\begin{theorem}[\cite{bounded}]The pair of operators $c^{\pm}_\rf(t)$ defined in \eqref{eq-e3} are the covariances of a pure, regular Hadamard state.
\end{theorem}

We stress that in general $c^\pm_\rf(t)$ are not `canonical' nor `distinguished', because they depend on the choice of the reference time $t_0$ and on the precise choice of the operators $b^\pm(t)$ (to which one can always add suitable regularizing terms). On the other hand, in Sect. \ref{secscat} we will construct covariances $c^\pm_{\inn}(t)$ and $c^\pm_{\out}(t)$ of the distinguished \emph{in} and \emph{out} states, and the operators $c^\pm_{\rf}(t)$ will play an important role in the proof of their Hadamard property: a suitable sufficient condition for that is in fact that
\beq
c^\pm_{\sca}(t)-c_{\rf}^\pm(t)\in \cW^{-\infty}(\Sigma)\otimes B(\cc^2)
\eeq
for some (and hence all) $t\in\rr$.

\subsection{Further estimates in scattering settings}\init\label{secscat}
 In what follows we give a refinement of the constructions in Sect. \ref{sec2} for the model Klein-Gordon equation in a scattering situation, corresponding to a situation when the metric $\altg$, resp. the potential $\altV$ converge to ultra-static metrics $\altg_{\rm out/in}= -dt^{2}+ \alth_{\rm out/in, ij}(\rx)d\rx^{i}d\rx^{j}$, resp. to time-independent potentials $\altV_{\outin}$ as $t\to \pm\infty$.
We start by fixing two classes of assumptions on the model Klein-Gordon equation \eqref{eq:modelP}.
\medskip

We  will often abbreviate  the classes $\Psi_{\td}^{m, \delta}(\rr;\Sigma)$ (introduced in Subsect. \ref{pdosec}-\ref{ss:auxr}) by $\Psi_{\td}^{m, \delta}$. We make the following assumption:
\[
(\Htd) \ \ \beal
a(t, \rx, D_{\rx})= a_{\outin}(\rx, D_{\rx})+ \Psi_{\td}^{2, -\delta}(\rr; \Sigma)  \hbox{ on }\rr^{\pm}\times \Sigma,\ \delta>0,\\[2mm]
r(t)\in \Psi_{\td}^{0, -1-\delta}(\rr; \Sigma),\\[2mm]
a_{\outin}(\rx, D_{\rx})\in \Psi^{2}(\Sigma) \hbox{\ elliptic}, \ a_{\outin}(\rx, D_{\rx})= a_{\outin}(\rx, D_{\rx})^{*}\geq C_{\infty}>0.
\eeal  
\]

The assumption $\delta>0$ corresponds to a \emph{long-range} setup (as opposed to the more narrow \emph{short-range} case when $\delta>1$).

\medskip

Below, we give estimates on the solution of the Riccati equation, taking now into account the decay in time that follows from  $(\Htd)$. To simplify notation we simply write $a_{1}(t)=a_{2}(t)+ \Psi_{\td}^{m, \delta}(\rr^{\pm}; \Sigma)$ when $a_{1}(t)=a_{2}(t)+ \Psi_{\td}^{m, \delta}(\rr; \Sigma)$ in $\rr^{\pm}\times \Sigma$. We also abbreviate $\Psi_{\td}^{m, \delta}(\rr^{\pm}; \Sigma)$ by $\Psi_{\td}^{m, \delta}$ when it is clear from the context whether the future or past case is meant.

From hypothesis $(\td)$ we deduce that there exists $c(t)\in \coinf(\rr)$ such that $a(t)+ c(t)\one\sim a_{\outin}$, uniformly in $t\in\rr^{\pm}$. By functional calculus we can find $\varphi\in \coinf(\rr)$ such that $a(t)+ \varphi(\frac{a(t)}{c(t)})\sim a_{\outin}$, uniformly in $t\in \rr^{\pm}$.  The error term  $ \varphi(\frac{a(t)}{c(t)})$ belongs to $\coinf(\rr; \cW^{-\infty}(\Sigma))$. 

We can hence replace $a(t)$ by $a(t)+ \varphi(\frac{a(t)}{c(t)})$ in the Riccati  equation \eqref{enp.11c} and assume that
\[
a(t)\sim  a_{\outin}\hbox{ uniformly in }t\in\rr^{\pm}.
\]
If $\epsilon_{\outin}\defeq  a_\outin^{\12}$, then from Prop.  \ref{l5.1} we deduce that if $(\td)$ holds then
\begin{equation}
\label{e5.1}
\epsilon(t)\defeq  a(t)^{\12}=  \epsilon_\outin+ \Psi^{1, -\delta}_{\td}.
\end{equation}

\begin{proposition}
 \label{p5.1}There exists $b(t)= \epsilon(t)+ \Psi_{\td}^{0, -1-\delta}(\rr; \Sigma)= \epsilon_{\outin}+ \Psi^{1,-\delta}_{\td}(\rr^{\pm}; \Sigma)$ that solves 
\[
 \i\p_{t}b- b^{2}+ a + \i rb\in \Psi_{\td}^{-\infty, -1-\delta}(\rr; \Sigma).
 \]
 \end{proposition}
 The proof is given in Appendix \ref{to1}.

\begin{proposition}\label{propoesti}
 Assume $(\td)$ and let $r_{b}^{\pm}$ be defined in \eqref{eq:rbpm} and $r_{-\infty}^{\pm}$ in \eqref{eq-ric}. Then
 \[
 r_{b}^{\pm}\in \Psi^{0, -1- \delta}_{\td}(\rr; \Sigma), \ \ r_{-\infty}^{\pm}\in \Psi^{-\infty, -1- \delta}_{\td}(\rr; \Sigma).
 \]
 \end{proposition}
 The proof is given in Appendix \ref{to2}.

\section{The $\outin$ states on asymptotically static spacetimes}\init\label{inout}
\subsection{Assumptions}\label{ss:asast} In what follows we introduce a class of asymptotically static spacetimes on which we will construct the \emph{in} and \emph{out} states and prove their Hadamard property. One of the key ingredients is the reduction to a model Klein-Gordon operator that satisfies the assumptions $(\Htd)$ considered in Subsect. \ref{secscat}.

We will use the framework of manifolds and diffeomorphisms of bounded geometry introduced in Defs. \ref{defp0.2}, \ref{defdeboun}.

We fix a  $d-$dimensional manifold $\Sigma$ equipped with a reference Riemannian metric $\altk$ such that $(\Sigma, \altk)$ is of bounded geometry, and consider $M= \rr_{t}\times\Sigma_{\ry}$, setting $y= (t, \ry)$, $n= 1+d$.  We equip $M$ with a Lorentzian metric $\altg$ of the form
\begin{equation}
\label{e10.1}
\altg= - \altc^{2}(y)dt^{2}+ (d\ry^{i}+ \altb^{i}(y)dt)\alth_{ij}(y)(d\ry^{j}+ \altb^{j}(y)dt),
\end{equation} 
where we assume:
\[
(\bg)\ \ \beal
\alth_{ij}\in \cinfb(\rr; \BT^{0}_{2}(\Sigma, \altk)), \  \alth_{ij}^{-1}\in \cinfb(\rr; \BT^{2}_{0}(\Sigma, \altk)),\\[2mm]
\altb\in \cinfb(\rr; \BT^{1}_{0}(\Sigma, \altk)), \\[2mm]
\altc, \ \altc^{-1}\in \cinfb(\rr; \BT^{0}_{0}(\Sigma, \altk)).
\eeal
\]
We recall that  $\tilde{t}\in \cinf(M)$ is called a  \emph{time function} if  $\nabla \tilde{t}$ is a timelike vector field. It is called a \emph{Cauchy time function} if its level sets are Cauchy hypersurfaces.
By \cite[Thm. 2.1]{CC} we know that $(M, \altg)$ is globally hyperbolic and  $t$ is  a Cauchy time function.

We will consider the Klein-Gordon operator on $(M, \altg)$:
\begin{equation}
\label{e10.2}
P= - \Box_{\altg}+ \altV,
\end{equation}
with $\altV\in \cinfb(\rr; \BT^{0}_{0}(\Sigma, \altk))$ a smooth real-valued function.
We consider two static metrics 
\[
\altg_{\outin}= -\altc^{2}_{\outin}(\ry)dt^{2}+ \alth_{\outin}(\ry)d\ry^{2}
\] 
and time-independent potentials $\altV_{\outin}$ and assume the following conditions:
\[
(\ast)\ \ \beal
\alth(y)- \alth_{\outin}(\ry)\in S^{-\mu}(\rr^{\pm}; \BT^{0}_{2}(\Sigma, \altk)), \\[2mm]
\altb(y)\in S^{-\mu'}(\rr; \BT^{1}_{0}(\Sigma, \altk)), \\[2mm]
\altc(y)- \altc_{\outin}(\ry)\in S^{-\mu}(\rr^{\pm}; \BT^{0}_{0}(\Sigma, \altk)),\\[2mm]
\altV(y)- \altV_{\outin}(\ry)\in S^{-\mu}(\rr^{\pm}; \BT^{0}_{0}(\Sigma, \altk)),
\eeal
\]
\[
(\pos)\ \ \frac{n-2}{4(n-1)}(\altR_{\altc_{\outin}^{-2}\alth_{\outin}}- \altc_{\outin}^{2}\altR_{\altg_\outin})+ \altc_{\outin}^{2}\altV_{\outin}\geq \altm^{2},
\]
for some $\mu>0$, $\mu'>1$ and $\altm>0$.  Above, $\altR_{\altg}$, resp. $\altR_{\alth}$ denotes the scalar curvature of $\altg$, resp. $\alth$.

Condition $(\ast)$ means that $\altg$, resp. $\altV$ are asymptotic to the static metrics $\altg_{\outin}$, resp. to the time-independent potentials  $\altV_{\outin}$ as $t\to \pm \infty$. Condition $(\pos)$ means that the asymptotic Klein-Gordon operators $\pe_{t}^{2}+ a_{\outin}(\rx, \pe_{\rx})$ introduced in Lemma \ref{l10.2} below   are {\em massive}. 

 It follows from $(\bg)$ that $\alth_{\outin}\in \BT^{0}_{2}(\Sigma, \altk)$, $\alth_{\outin}^{-1}\in \BT^{2}_{0}(\Sigma, \altk)$, and that $\altV_{\outin}, \altV_{\outin}^{\,-1}\in \BT^{0}_{0}(\Sigma, \altk)$.
\subsection{Reduction to the model case}\label{s10.1}
In this subsection we perform the reduction of the Klein-Gordon operator $P$ to the model case considered in Sect. \ref{secscat}. 
We start with the well-known orthogonal decomposition of $\altg$ associated with the time function $t$. Namely, we set
\[
v\defeq  \dfrac{\altg^{-1}dt}{dt\cdot \altg^{-1}dt}=\pe_{t}+ \altb^{i}\pe_{\ry^{i}},
\]
which using  $(\bg)$ is a complete vector field. Furthermore, we denote by $\phi_{t}$ its flow, so that 
\[
\phi_{t}(\rx)= (t, \ry(t,0, \rx)),\ t\in \rr, \ \rx\in \Sigma,
\]
where $\ry(t, s, \cdot)$ is the flow of the time-dependent vector field $b$ on $\Sigma$.
We also set
\beq\label{e10.0}
\chi: \rr\times \Sigma\ni(t, \rx)\mapsto (t, \ry(t, 0, \rx))\in \rr\times \Sigma.
\eeq

\begin{lemma}\label{l10.1}Assume $(\bg)$, $(\ast)$. Then
\[
\altgh:= \chi^{*}\altg= -  \altch^{2}(t, \rx)dt^{2}+ \hat  \alth(t, \rx)dy^{2}, \ \ \chi^{*}\altV=  \altVh, 
\]
where:
\[
\begin{array}{l}
 \altch,  \altch^{-1},  \altVh\in \cinfb(\rr; \BT^{0}_{0}(\Sigma, \altk)),\\[2mm]
 \hat \alth\in \cinfb(\rr; \BT^{0}_{2}(\Sigma, \altk)), \  \hat \alth^{-1}\in \cinfb(\rr; \BT^{2}_{0}(\Sigma, \altk)).
\end{array}
\]
Moreover there exist bounded diffeomorphisms $\ry_{\outin}$ of $(\Sigma, \altk)$ such that if:
\[
\begin{array}{l}
 \hat \alth_{\outin}\defeq \ry_{\outin}^{*}\alth_{\outin},\\[2mm]
   \altch_{\outin}\defeq  \ry_{\outin}^{*}\altc_{\outin}, \  \altVh_{\outin}\defeq  \ry_{\outin}^{*}\altV_{\outin},
\end{array}
\]
then we have:
\[
\begin{array}{l}
 \hat \alth_{\outin}\in \BT^{0}_{2}(\Sigma, \altk), \  \hat \alth_{\outin}^{-1}\BT^{2}_{0}(\Sigma, \altk),\\[2mm]
 \altch_{\outin},  \altch_{\outin}^{-1},  \altVh_{\outin}\in \BT^{0}_{0}(\Sigma, \altk),
\end{array}
\]
and furthermore,
\[
\begin{array}{rl}
 \hat \alth- \hat  \alth_{\outin}\in S^{-\min(1-\mu', \mu)}(\rr^{\pm}, \BT^{0}_{2}(\Sigma, \altk)), \\[2mm]
 \altch- \altch_{\outin}\in S^{-\min(1-\mu', \mu)}(\rr^{\pm}, \BT^{0}_{0}(\Sigma, \altk)),\\[2mm]
 \altVh-  \altVh_{\outin}\in S^{-\mu}(\rr^{\pm}, \BT^{0}_{0}(\Sigma, \altk)).
\end{array}
\]
\end{lemma}
Lemma \ref{l10.1} is proved in Appendix \ref{apoti}.

Writing $P$ as $-\Box_{\altg}+  \frac{n-2}{4(n-1)}\altR_{\altg}+ \altW\,$ for $\altW= \altV-  \frac{n-2}{4(n-1)}\altR_{\altg}$, and using the conformal invariance of $-\Box_{\altg}+  \frac{n-2}{4(n-1)}\altR_{\altg}$ and the estimates in Lemma \ref{l10.1},
we obtain the following result, which completes the reduction to the model case. 

If $P\in {\rm Diff}(\rr\times \Sigma)$ we denote by $\chi^{*}P$ the pullback of $P$ by $\chi$ defined by  $(\chi^{*}P)u\circ \chi=(Pu)\circ \chi$.
\begin{lemma}\label{l10.2}
Assume $(\bg)$, $(\ast)$, $(\pos)$ and consider the Klein-Gordon operator $P$  in \eqref{e10.2}. Let  $\hat{\alth}, \altch,\altVh$ be as in Lemma \ref{l10.1} and set:
\[
\hat P=  \chi^{*}P, \ \ \tilde{P}=  \altch^{1-n/2}\hat P \altch^{1+ n/2}, \ \ \altgt= \altch^{-2}\altgh, \ \ \tilde{\alth}=  \altch^{-2}\hat{\alth}.
\]
Then 
\[
\tilde{P}= \pe_{t}^{2}+r(t, \rx)\pe_{t}+ a(t, \rx, \pe_{\rx}),
\]
for
\[
\begin{array}{l}
a(t, \rx, \pe_{\rx})=- \Delta_{\tilde{\alth}_{t}}+\altVt, \ \ r=  |\tilde{\alth}_{t}|^{-\12} \p_{t}|\tilde{\alth}_{t}|^{\12},\\[2mm]
\altVt= \frac{n-2}{4(n-1)}(R_{\altgt}- \altch^{2}R_{\altgh})+ \altch^{2}\altVh.
\end{array}
\]
Moreover $a, r$ satisfy $(\Htd)$ with $\delta= \min (\mu, \mu'-1)$ and
\[
a_{\outin}(\rx, \pe_{\rx})= - \Delta_{\tilde{\alth}_{\outin}}+\altVt_{\outin}(\rx),
\]
where
\[
\altVt_{\outin}= \left(\frac{n-2}{4(n-1)}(\altR_{\altc^{-2}_{\outin}h_{\outin}}- \altc^{2}_{\outin }\altR_{\altg_{\outin}})+ \altc_{\outin}^{2}\altV_{\outin}\right)\circ \ry_{\outin}.
\]\end{lemma}
Note that condition $(\pos)$ simply means that $\altVt_{\outin}\geq \altm^{2}>0$.
\subsection{Cauchy evolutions}\label{cauchycauchy}
In this subsection we relate the Cauchy evolutions of $P$ and  of the model Klein-Gordon operator $\tilde{P}$.

The trace operator for $P$ associated to the time function $t$ is given by:
\beq\label{defdetrace}
\varrho_{t}\phi= \col{u(t, \cdot)}{\i^{-1}n\cdot\nabla \phi(t, \cdot)},
\eeq
where $n$ is the future directed unit normal to $\Sigma_{t}$. The corresponding trace operator for $\hat{P}= \chi^{*}P$ is:
\[
\hat{\varrho}_{t}\phi= \col{\phi(t, \cdot)}{ \i^{-1}\altch^{-1}\p_{t}\phi(t, \cdot)},
\]
so that denoting $\chi^{*}\phi= \phi\circ \chi$, we have:
\[
\hat{\varrho}_{t}\chi^{*}\phi= \chi_{t}^{*}\varrho_{t}\phi \, \hbox{ for } \chi_{t}^{*}\col{u_{0}}{u_{1}}= \col{u_{0}\circ \chi_{t}}{u_{1}\circ \chi_{t}},
\]
and $\chi_{t}(\rx)= \ry(t, 0, \rx)$, see \eqref{e10.0}. Finally the trace operator for $\tilde{P}$ as in Lemma \ref{l10.2} is
\[
\tilde{\varrho}_{t}\phi= \col{\phi(t, \cdot)}{\i^{-1}\phi(t, \cdot)}
\]
so that if $\tilde{\phi}= \altch^{n/2-1}\phi$  is the conformal transformation in Lemma \ref{l10.2}, we have:
\[
\tilde{\varrho}_{t}\tilde{\phi}= R(t)\hat{\varrho}_{t}\phi,\hbox{ for }R(t)= \altch^{n/2-1}\mat{1}{0}{-\i(n/2-1)\p_{t}\ln (\altch)}{1}.
\]
Let us denote by $\cU(t,s)$ the Cauchy evolution for $P$ associated to $\varrho_{t}$ and by $\cU^{\adg}(t,s)$ the almost diagonal Cauchy evolution introduced in 
Subsect. \ref{ss:iad} for the model Klein-Gordon operator $\tilde P$.  The following lemma  follows from the above computations and \eqref{intolo}.
\begin{lemma}\label{defdeZ}
Let $Z(t)\defeq (\chi_{t}^{*})^{-1} R(t)T(t)$, where $T(t)$ is defined in  \eqref{eq:defT}. Then
\beq\label{e10.5}
\cU(t,s)= Z(t)\cU^{\adg}(t,s)Z^{-1}(s).
\eeq
\end{lemma}

We have a similar reduction for  the asymptotic Klein-Gordon operators:
\[
P_{\outin}= - \Box_{\altg_{\outin}}+\frac{n-2}{4(n-1)}\altR_{g_{\outin}}+ \altV_{\outin},
\]
for $\altg_{\outin}= -\altc^{2}_{\outin}(\ry)dt^{2}+ \alth_{\outin}(\ry)d\ry^{2}$, where $\alth_{\outin}, \altc_{\outin}, \altV_{\outin}$ were introduced in $(\ast)$. The associated trace operator is
\[
\varrho_{t,\outin}\phi= \col{\phi(t, \cdot)}{\i^{-1}\altc_{\outin}^{-1}\p_{t}\phi(t, \cdot)}.
\]
We also set
\[
\chi_{\outin}^{*}\col{u_{0}}{u_{1}}= \col{u_{0}\circ \ry_{\outin}}{u_{1}\circ \ry_{\outin}},\ \ R_{\outin}=\altch_{\outin}^{(d-1)/2}\one,
\]
and for $\epsilon_{\outin}= a_{\outin}^{\12}$:
\[
T_{\outin}= (\i \sqrt{2})^{-1}\mat{\epsilon_{\outin}^{-\12}}{-\epsilon_{\outin}^{-\12}}{\epsilon_{\outin}^{\12}}{\epsilon_{\outin}^{\12}}, \ \ Z_{\outin}= (\chi_{\outin}^{*})^{-1}R_{\outin}T_{\outin},
\]
so that the Cauchy evolution of $P_{\outin}$ is given by
\begin{equation}
\label{e10.6}
\cU_{\outin}(t,s)= Z_{\outin}\circ \cU_{\outin}^{\adg}(t,s)\circ Z_{\outin}^{-1}, 
\end{equation}
where $\cU_{\outin}^{\adg}$ stands for the evolution generated by
\begin{equation}
\label{e10.7}
 H^{\adg}_{\outin}= \mat{\epsilon_{\outin}}{0}{0}{\epsilon_{\outin}}.
\end{equation}
The following fact will be needed in the sequel.
\begin{lemma}\label{l10.3}We have:
\[
Z^{-1}(t)Z_{\outin}-\one, \ Z_{\outin}^{-1}Z(t)-\one \to 0\hbox{ in }B(L^{2}(\Sigma)\otimes \cc^{2})\hbox{ as }t\to \pm \infty.
\]
\end{lemma}
\proof
From Prop. \ref{p5.1} we obtain that $T^{-1}_{\outin}T(t)-\one$ tends to $0$ in norm as $t\to \pm \infty$.
By Lemma \ref{l10.1}, $R(t)$ tends to $R_{\outin}$ in norm. Finally, from the proof of Lemma \ref{l10.1}, see in particular \eqref{e10.4}, we obtain that 
$(\chi_{\outin}^{*})^{-1}\chi_{t}^{*}$ tends  to $\one$ in norm. This implies the lemma. \qed
\subsection{Construction of Hadamard states by scattering theory}
In this subsection we construct the \emph{out}/\emph{in} states $\omega_{\outin}$ for the Klein-Gordon operator $P$ and show that they are Hadamard states. We assume hypotheses $(\bg)$, $(\ast)$, $(\pos)$.

By the positivity condition $(\pos)$, the asymptotic Klein-Gordon operators $P_{\outin}$ admit {\em vacuum states} (that is, {\em ground states} for the dynamics $\cU_\outin$) $\omega^{\rm vac}_\outin$.  In terms of $t=0$ Cauchy data  their covariances are the projections:
\[
c^{\pm, {\rm vac}}_{\outin}= Z_{\outin}\pi^{\pm}Z_{\outin}^{-1}, 
\]
where  $\pi^{\pm}$ are defined in \eqref{defdepiplusmoins}.  Clearly we have
\[
\cU_{\outin}(t,s)c^{\pm, {\rm vac}}_{\outin}\cU_{\outin}(s,t)= c^{\pm, {\rm vac}}_{\outin},
\]
i.e. $\omega^{\rm vac}_{\outin}$ are invariant under the asymptotic dynamics.
 For $t\in\rr$ we now consider the projections:
 \beq\label{eq:defcp}
\bea
c^{\pm,t}_{\outin}(0)&\defeq  \cU(0, t)c_{\outin}^{\pm, {\rm vac}}\cU(t,0)\\
&= \cU(0, t)\cU_{\outin}(t, 0)c_{\outin}^{\pm, {\rm vac}}\cU_{\outin}(0, t)\cU(t,0).
\eea
\eeq
By taking the $t\to\pm\infty$ limit of $c^{\pm,t}_{\outin}(0)$ we  obtain the time-$0$ covariances  $c^{\pm}_{\outin}(0)$  of  a state $\omega_\outin$ (for the Klein-Gordon operator $P$) that `equals $\omega^{\rm vac}_{\outin}$ asymptotically' at $t=\pm\infty$. The main new result that we prove is that  $\omega_\outin$ are Hadamard states.

  Before stating the theorem let us recall that the Sobolev spaces $H^{\sobo}(\Sigma)$ are naturally defined using the reference Riemannian metric $\altk$ on $\Sigma$. The {\em charge space} $H^{\12}(\Sigma)\oplus H^{-\12}(\Sigma)$ is the natural space of Cauchy data in connection with quantized Klein-Gordon fields. 

\begin{theorem}\label{thm.scat1}Assume hypotheses $(\bg)$, $(\ast)$, $(\pos)$. Then 
 \beq
\lim_{t\to \pm\infty}c^{\pm,t}_{\outin}(0)\eqdef c^{\pm}_{\outin}(0)=  c_{\rf}^{\pm}(0)+ \cW^{-\infty}(\Sigma), \hbox{ in }B(H^{\12}(\Sigma)\oplus H^{-\12}(\Sigma)),
\eeq
where $c^{\pm}_{\rm ref}(0)= Z(0)\pi^{\pm}Z^{-1}(0)$. The operators $c^{\pm}_{\outin}(0)$ are pairs of projections defining a pure state $\omega_{\outin}$ for the Klein-Gordon operator $P$. Moreover $\omega_{\outin}$ is a Hadamard state.
\end{theorem}
\proof  From \eqref{e10.5}, \eqref{e10.6}  we obtain:
\begin{equation}
\label{e5.3}\bea 
\cU_{\outin}(0, t)\cU(t, 0)&= Z_{\outin}(0)\cU^{\adg}_{\outin}(0, t)Z^{-1}_{\outin}Z(t)\cU^{\adg}(t,0)Z^{-1}(0),\\[2mm]
\cU(0, t)\cU_{\outin}(t, 0)&= Z(0)\cU^{\adg}(0, t)Z^{-1}(t)Z_{\outin}\cU^{\adg}_{\outin}(t,0)Z_{\outin}^{-1}.
\eea 
\end{equation}
 It follows that:
\begin{equation}
\label{e5.3c}
\bea
c^{\pm,t}_{\outin}(0)&=Z(0)\cU^{\adg}(0,t)Z^{-1}(t)Z_{\outin}\cU^{\adg}_{\outin}(t, 0)\\
&\phantom{=}\times\pi^{\pm}\cU^{\adg}_{\outin}(0, t)Z^{-1}_{\outin}Z(t)\cU^{\adg}(t, 0)Z^{-1}(0).
\eea
\end{equation}
Since $Z(0):L^{2}(\Sigma)\otimes \cc^{2}\to H^{\12}(\Sigma)\oplus H^{-\12}(\Sigma)$ is boundedly invertible it suffices to  show the existence of the limit
\[
\bea
d^{\pm}_{\outin}&= \lim_{t\to \pm\infty}\cU^{\adg}(0,t)Z^{-1}(t)Z_{\outin}\cU^{\adg}_{\outin}(t,0)\pi^{\pm}\\
&\phantom{\,=\lim_{t\to \pm\infty}}\times\cU^{\adg}_{\outin}(0, t)Z^{-1}_{\outin}Z(t)\cU^{\adg}(t, 0)
\eea 
\]
in $B(L^{2}(\Sigma)\otimes \cc^{2})$.

By Prop.  \ref{l.scat1} (1) below we know that $\cU^{\adg}(t,s)$, $\cU^{\adg}_{\outin}(t,s)$ are uniformly bounded in $B(L^{2}(\Sigma)\otimes \cc^{2})$. Hence using Lemma \ref{l10.3} we can replace $Z^{-1}(t)Z_{\outin}$ and $Z^{-1}_{\outin}Z(t)$ by $\one$ in the rhs of \eqref{e5.3c}, modulo an error of size $o(t^{0})$ in $B(L^{2}(\Sigma)\otimes \cc^{2})$, i.e.  we are reduced to prove the existence of the limit
\[
\bea 
d^{\pm}_{\outin}&\defeq \lim_{t\to \pm\infty}\cU^{\adg}(0,t)\cU^{\adg}_{\outin}(t, 0)\pi^+\cU^{\adg}_{\outin}(0, t)\cU^{\adg}(t, 0)\\[2mm]
&=\lim_{t\to \pm\infty}W_{\outin}(t)\pi^+ W_{\outin}^{-1}(t),
\eea 
\]
where $W_{\outin}(t)= \cU^{\adg}(0, t)\cU^{\adg}_{\outin}(t, 0)$.  By Prop. \ref{l.scat1} the limit exists in $B(L^{2}(\Sigma)\otimes \cc^{2})$ and equals $\pi^++ \cW^{-\infty}(\Sigma)$.  The limit operators $d^{\pm}_{\outin}$ are  projections as  norm limits of projections. It follows that
\beq\label{ebou}
c^{\pm}_{\outin}(0)= Z(0)d^{\pm}_{\outin}Z(0)^{-1}+ \cW^{-\infty}(\Sigma)= c^{+}_{\rf}(0) + \cW^{-\infty}(\Sigma)
\eeq
is a projection.  The conditions \eqref{eq:secondcondlam0}, \eqref{eq:secondcondlam} are satisfied by $c^{\pm}_{\outin}(0)$ since they are satisfied by $c^{\pm,t}_{\outin}(0)$ for each finite $t$.  Therefore $c^{\pm}_{\outin}$ are the covariances of two pure states $\omega_{\outin}$ for $P$.  Finally as in \cite{bounded} we obtain from \eqref{ebou} that $\omega_{\outin}$ are Hadamard states. \qeds

In the proof of Thm. \ref{thm.scat1}, the crucial ingredient is the following proposition.

\begin{proposition}\label{l.scat1}
 Let $H^{\adg}(t),H^{\adg}_{\outin}$ be  as in \eqref{eq:defH}, \eqref{e10.7}. Then:
 \ben
 \item $\cU^{\adg}_{\outin}(t,s)$ and $\cU^{\adg}(t,s)$ are uniformly bounded in $B(H^{m}(\Sigma)\otimes \cc^{2})$, for all $m\in \rr.$
 \item Let $W_{\outin}(t)= \cU^{\adg}(0, t)\cU^{\adg}_{\outin}(t, 0)$. Then  
 \[
\lim_{t\to +\infty} W_{\outin}(t)\pi^+ W_{\outin}(t)^{-1}= \pi^+ + \cW^{-\infty}(\Sigma)\otimes L(\cc^{2}), \hbox{ in }B(L^{2}(\Sigma)\otimes \cc^{2}).
 \]
  \een
\end{proposition}
\proof  {\it Proof of (1)}: we can assume without loss of generality that $s=0$.  The statement for $\cU^{\adg}_{\outin}(t,0)$ is obvious since $H^{\adg}_{\outin}= \mat{\epsilon_{\outin}}{0}{0}{-\epsilon_{\outin}}$. Let us prove it for $\cU^{\adg}(t,0)$. We have: 
\beq\label{e5.5b}
\bea 
H^{\adg}(t)&=  \mat{-b^{-}(t)+ \i r_{b}^- (t)}{0}{0}{-b^{+}(t)+\i r_{b}^+(t)}+ \Psi_{\td}^{-\infty, -1-\delta}(\rr; \Sigma)\otimes B(\cc^{2})\\[2mm]
&=\mat{\epsilon(t)}{0}{0}{-\epsilon(t)}+  \Psi_{\td}^{0, -1-\delta}(\rr; \Sigma)\otimes B(\cc^{2}),
\eea 
\eeq
 by  Props. \ref{p5.1}, \ref{propoesti}. Since $\epsilon(t)$ is selfadjoint, this implies that $\cU^{\adg}(t,0)$  is uniformly bounded in $B(L^{2}(\Sigma))$, which proves (1) for $m=0$.

We now note that $\| u\|_{H^{m}(\Sigma)}\sim \| \epsilon^{m}(t)u\|_{L^{2}(\Sigma)}$, uniformly for $t\in \rr$, since  $\epsilon(t)$ is elliptic uniformly for $t\in \rr$.  Therefore to prove (1) it suffices, using the uniform boundedness of $\cU^{\adg}(t,0)$ in $B(L^{2}(\Sigma))$, to show that
\beq\label{e5.10}
\cU^{\adg}(0,t)\left(\epsilon(t)^{m}\otimes \one_{\cc^{2}}\right)\cU^{\adg}(t,0) \left(\epsilon(0)^{-m}\otimes \one_{\cc^{2}}\right)\hbox { is uniformly bounded}
\eeq
in $B(L^{2}(\Sigma))$. We have by \eqref{e5.5b}: 
\[
\bea 
&\p_{t}\cU^{\adg}(0,t)\left(\epsilon(t)^{m}\otimes \one_{\cc^{2}}\right)\cU^{\adg}(t,0) \left(\epsilon(0)^{-m}\otimes \one_{\cc^{2}}\right)\\[2mm]
&= \cU^{\adg}(0,t)\left(\p_{t}\epsilon^{m}(t)\otimes\one_{\cc^{2}}+ \i [H^{\adg}(t), \epsilon^{m}(t)\otimes \one_{\cc^{2}}]\right)\cU^{\adg}(t,0)\left(\epsilon(0)^{-m}\otimes \one_{\cc^{2}}\right)\\[2mm]
&= \cU^{\adg}(0,t)\left(\p_{t}\epsilon^{m}(t)\otimes \one_{\cc^{2}}+ \i [H^{\adg}(t), \epsilon^{m}(t)\otimes \one_{\cc^{2}}]\right)\left(\epsilon(t)^{-m}\otimes\one\right)\cU^{\adg}(t,0)\\[2mm]
&\phantom{=}\times \cU^{\adg}(0,t)\left(\epsilon(t)^{m}\otimes \one_{\cc^{2}}\right)\cU^{\adg}(t,0)\left(\epsilon(0)^{-m}\otimes \one_{\cc^{2}}\right)\\[2mm]
&\eqdef  M(t) \cU^{\adg}(0,t)\left(\epsilon(t)^{m}\otimes \one_{\cc^{2}}\right)\cU^{\adg}(t,0)\left(\epsilon(0)^{-m}\otimes \one_{\cc^{2}}\right).
\eea 
\]
By  $(\Htd)$ and Prop.  \ref{l5.1} we  see that  $\p_{t}\epsilon^{m}(t)\in \Psi_{\td}^{m, -1- \delta}$, and by \eqref{e5.5b}   that $ [H^{\adg}(t), \epsilon^{m}(t)\otimes \one_{\cc^{2}}]\in \Psi_{\td}^{m, -1-\delta}$. Therefore 
$\|M(t)\|_{B(L^{2}(\Sigma)\otimes \cc^{2})}\in O(\langle t\rangle^{-1- \delta})$. Hence, setting
\[
f(t)\defeq\|\cU^{\adg}(0,t)\left(\epsilon(t)^{m}\otimes \one_{\cc^{2}}\right)\cU^{\adg}(t,0) \left(\epsilon(0)^{-m}\otimes \one_{\cc^{2}}\right)\|_{B(L^{2}(\Sigma))},
\]
we have $f(0)=1$, $|\p_{t}f(t)|\in O(\langle t\rangle^{-1- \delta})f(t)$. If $f(t)\neq +\infty$  for each $t$, an application of 
 Gronwall's inequality  would immediately imply \eqref{e5.10}.  If $m\leq 0$ the  use of Gronwall's inequality is justified by applying the above time dependent operator to a vector $u\in H^{m}(\Sigma)$. If $m>0$ we replace the unbounded operator $A= \epsilon(t)\otimes \one_{\cc^{2}}$ by the bounded operator
 $A_{\delta}= A(1+ \i \delta A)$, for $\delta>0$. For the corresponding function $f_{\delta}(t)$ we obtain that
  $f_{\delta}(0)\leq 1$, $|\p_{t}f_{\delta}(t)|\in O(\langle t\rangle^{-1- \delta})f_{\delta}(t)$ uniformly for $0<\delta\leq 1$.
  Then \eqref{e5.10} follows using that $\|A^{m}u\|= \sup_{0<\delta\leq 1}\| A_{\delta}^{m}u\|$.

{\it Proof of  (2)}: note first that $[\pi^+, A]=0$ for any diagonal operator $A$. Therefore:
 \[
W_{\outin}(t)\pi^+ W_{\outin}(t)^{-1}= \cU(0,t)\pi^+ \cU(t,0),
\]
and by \eqref{e5.5b}
\beq\label{eq:rin}
\begin{array}{l}
\p_{t}(W_{\outin}(t)\pi^+ W_{\outin}(t)^{-1})= - \i \cU(0,t)[H^{\adg}(t), \pi^+]\cU(t,0)\\[2mm]
= \cU(0,t)[R_{-\infty}(t), \pi^+]\cU(t,0), \ R_{-\infty}\in \Psi_{\td}^{-\infty, -1- \delta}(\rr; \Sigma)\otimes B(\cc^{2}).
\end{array}
\eeq
By (1), this implies that $\p_{t}(W_{\outin}(t)\pi^+ W_{\outin}(t)^{-1})\in \Psi_{\td}^{-\infty, -1- \delta}(\rr; \Sigma)\otimes B(\cc^{2})$, hence:
\[
\bea
&\lim_{t\to +\infty}W_{\outin}(t)\pi^+ W_{\outin}(t)\\ &= \pi^++ \int_{0}^{+\infty}\p_{t}(W_{\outin}(t)\pi^+ W_{\outin}(t)^{-1}) dt \hbox{\, in }B(L^{2}(\Sigma)\otimes \cc^{2}).
\eea
\]
The integral term belongs to $\cW^{-\infty}(\Sigma)$. \qeds

\appendix
\section{}\init\label{secapp1}
\subsection{Proof of Prop. \ref{l5.1}}\label{ssecap1}
To prove Prop. \ref{l5.1} we first need an auxiliary lemma about parameter-dependent pseudo\-differential calculus.

 We start by introducing parameter dependent versions of the spaces  $\Psi^{m}(\Sigma)$, $S^{0}(\rr; \Psi^{m}(\Sigma))$.
 
We define the  symbol classes $\widetilde{S}^{m}(T^{*}\Sigma)$  for $m\in \rr$ as the space of functions $c(\rx, \spexi, \lambda)\in \cinf(T^{*}\Sigma\times \rr)$ such that:
\[
\p_{\lambda}^{\gamma}\p_{\rx}^{\alpha}\p_{\spexi}^{\beta}c(\rx, \spexi, \lambda)\in O(\langle \spexi\rangle + \langle \lambda\rangle)^{m-|\beta|- \gamma}, \ \alpha, \beta\in \nn^{d}, \ \gamma\in \nn,
\]
as usual understood after fixing a good chart cover and good chart diffeomorphisms, with uniformity of the constants with respect to the element of the cover.  The standard example of such a symbol is $c(\rx, \spexi, \lambda)= (a(\rx, \spexi)+\langle \lambda\rangle^{m})$, for $a\in S^{m}(T^{*}\Sigma)$ elliptic and positive.
 
The subspaces of symbols poly-homogeneous in $(\spexi, \lambda)$ are denoted by $\widetilde{S}^{m}_{\rm ph}(T^{*}\Sigma)$.
We define  $\widetilde{\cW}^{-\infty}(\Sigma)$ as the set of smooth maps $\rr\ni \lambda\mapsto a(\lambda)\in \cW^{-\infty}(\Sigma)$ such that:
 \[
\| \p_{\lambda}^{\gamma}a(\lambda)\|_{B(H^{-m}(\Sigma), H^{m}(\Sigma))}\in O(\langle \lambda\rangle^{-n}), \ \forall m, n,\gamma\in \nn,
 \]
 and we set
 \[
 \widetilde{\Psi}^{m}(\Sigma)\defeq  \Op(\widetilde{S}_{\rm ph}^{m}(T^{*}\Sigma)) + \widetilde{\cW}^{-\infty}(\Sigma).
 \]
We  also define the time-dependent versions:
\[S^{0}(\rr; \widetilde{S}^{m}_{({\rm ph})}(T^{*}\Sigma)),  \ S^{0}(\rr; \widetilde{\cW}^{-\infty}(\Sigma)), \ S^{0}(\rr; \widetilde{\Psi}^{m}(\Sigma)),
\]
in analogy with Subsect. \ref{symbolo}. For example $c(t, \rx, \spexi, \lambda)\in S^{0}(\rr; \widetilde{S}^{m}(T^{*}\Sigma))$ if
\[
\p_{t}^n\p_{\lambda}^{\gamma}\p_{\rx}^{\alpha}\p_{\spexi}^{\beta}c(t,\rx, \spexi, \lambda)\in O(\langle t\rangle^{-n}(\langle \spexi\rangle + \langle \lambda\rangle)^{m-|\beta|- \gamma}), \ \alpha, \beta\in \nn^{d}, \ \gamma, n\in \nn.
\]

\begin{lemma}\label{lemomo}
 Let $a(t)\in S^{0}(\rr; \Psi^{2}(\Sigma))$ such that $a(t)$ is elliptic, selfadjoint on $L^{2}(\Sigma)$ with $a(t)\geq c_{0}\one$,  $c_{0}>0$. Then $(a(t)+ \lambda^{2})^{-1}\in S^{0}(\rr; \widetilde{\Psi}^{-2}(\Sigma))$.
 \end{lemma}
\proof  The proof is based on a reduction to the situation without   the parameter $\lambda$. 
We first present the argument in the time-independent case.

{\it Step 1}. Let us denote by $l\in \rr$ the dual variable to $\lambda$. We consider the manifold of bounded geometry $\Sigma_{\rx}\times \rr_{l}$ equipped with the metric $\alth_{ij}(\rx)d\rx^{i}d\rx^{j}+ dl^{2}$. As good chart covering we can take $\widetilde{U}_{i}= U_{i}\times \rr$, $\widetilde{\psi}_{i}(\rx, l)= (\psi_{i}(\rx), l)$ where $\{U_{i}, \psi_{i}\}_{i\in \nn}$ is a good chart covering for $(\Sigma, \alth)$.  A subordinate good partition of unity is $\widetilde{\chi}_{i}(\rx, l)= \chi_{i}(\rx)$.
 
 The classes $S^{m}_{\rm ph}(T^{*}(\Sigma\times \rr))$ are then defined as in Subsect. \ref{symbolo}.
 Denoting ${\rm ad}_{A}B\defeq [A, B]$, one sets and one sets:
 \beq\label{comutoto}
 \cW^{-\infty}(\Sigma\times \rr)= \{A\ : {\rm ad}^{\gamma}_{l}A\in B(H^{-m}(\Sigma\times \rr), H^{m}(\Sigma\times\rr)), m, \gamma\in \nn.\},
 \eeq
 This choice of the ideal $\cW^{-\infty}(\Sigma\times \rr)$ is dictated by the definition of $\tilde{\cW}^{-\infty}(\Sigma)$. We set  then  $\Psi^{m}(\Sigma\times \rr)= \widetilde{\Op}(S^{m}_{\rm ph}(T^{*}(\Sigma\times \rr)))+ \cW^{-\infty}(\Sigma\times \rr)$,
 where $\widetilde{\Op}$ is defined as in Subsect. \ref{pdosec} with $\Sigma$ replaced by $\Sigma\times \rr$. Note that because of our choice of the chart covering $\widetilde{\Op}$ is the usual Weyl quantization w.r.t. the $(l, \lambda)$ variables.
 
 {\it Step 2}. In step 2 we describe the link between $\Psi^{m}(\Sigma\times \rr)$ and $\tilde{\Psi}^{m}(\Sigma)$.
  We note that
 \[
 \widetilde{S}^{m}_{\rm ph}(T^{*}\Sigma)= \{c\in S^{m}_{\rm ph}(T^{*}(\Sigma\times \rr)): \ \p_{l}c=0\},
 \]
 and denoting by $T_{l}$ the group of translations in $l$ we have
 \[
 [T_{l}, \widetilde{\Op}(c)]=0, \forall\  l\in \rr \ \Leftrightarrow \ c\in \widetilde{S}_{\rm ph}^{m}(T^{*}\Sigma). 
 \]
Equivalently, if $\mathcal{F}$ is the Fourier transform in $l$ we have
 \beq\label{e.deco1}
 \bea
 &c\in S^{m}_{\rm ph}(T^{*}(\Sigma\times\rr)), \ [T_{l}, \widetilde{\Op}(c)]=0\\[2mm]
  \Leftrightarrow & \ \mathcal{F}\widetilde{\Op}(c)\mathcal{F}^{-1}= \int^{\oplus}_{\rr}\Op(c(\lambda))d\lambda, \hbox{ for } c(\lambda)\in \widetilde{S}^{m}_{\rm ph}(T^{*}\Sigma).
 \eea
  \eeq
 
 Let now $w\in \cW^{-\infty}(\Sigma\times \rr)$ with $[w, T_{l}]=0$.  We have:
 \beq\label{four}
 \mathcal{F}w\mathcal{F}^{-1}= \int^{\oplus}_{\rr} w(\lambda)d\lambda.
 \eeq
 Since ${\rm ad}^{\gamma}_{l}w\in \bigcap_{m\in \nn}B(H^{-m}(\Sigma\times \rr), H^{m}(\Sigma\times\rr))$ we obtain that:
 \[
 \int_{\rr}\langle \lambda\rangle^{n}\| \p^{\gamma}_{\lambda}w(\lambda)u(\lambda)\|^{2}_{H^{p}(\Sigma)}d\lambda\leq C_{n, p}\int_{\rr}\langle \lambda\rangle^{-n}\| u(\lambda)\|^{2}_{H^{-p}(\Sigma)}d\lambda, \ \forall \gamma,n, p\in \nn,
 \]
 or equivalently
 \[
 \int^{\oplus}_{\rr}\langle \lambda\rangle^{n}(-\Delta_{\alth}+1)^{p/2}\p_{\lambda}^{\gamma}w(\lambda)(-\Delta_{\alth}+1)^{p/2}d\lambda\in B(L^{2}(\Sigma\times \rr)).
 \]
 By Sobolev's embedding theorem this  implies that
 \[
 \|\p_{\lambda}^{\gamma}w(\lambda)\|_{B(H^{-p/2}(\Sigma), H^{p/2}(\Sigma))}\in O(\langle \lambda\rangle^{-n})\ \forall \gamma,n, p\in \nn,
 \]
  hence $
 w(\lambda)\in \widetilde{\cW}^{-\infty}(\Sigma)$.
 Conversely, if $w(\lambda)\in  \widetilde{\cW}^{-\infty}(\Sigma)$  it is immediate that  $w$ defined by \eqref{four}
 belongs to $\cW^{-\infty}(\Sigma\times \rr)$. Hence we have shown
 \begin{equation}
 \label{e.deco2}
 \bea
 & w\in \cW^{-\infty}(\Sigma\times \rr), \ [w, T_{l}]=0\\[2mm]
 \Leftrightarrow& \ \mathcal{F}w\mathcal{F}^{-1}= \int^{\oplus}_{\rr}w(\lambda)d\lambda, \hbox{ for }  w(\lambda)\in \widetilde{\cW}^{-\infty}(\Sigma). 
 \eea
  \end{equation}
 Let us now consider the time-dependent situation. If we  define the  time-dependent classes  $\cinfb(\rr; \widetilde{S}^{m}(T^{*}\Sigma))$, $\cinfb(\rr; \widetilde{\cW}^{-\infty}(\Sigma))$ and $\cinfb(\rr; \widetilde{\Psi}^{m}(\Sigma))$ in the obvious way, then 
 \begin{equation}
 \label{e.deco3}
 \bea
 &c\in \cinfb(\rr; S^{m}_{\rm ph}(T^{*}(\Sigma\times \rr))), \  [T_{l}, \widetilde{\Op}(c)(t)]=0\\[2mm]
 \Leftrightarrow& \ \mathcal{F}\widetilde{\Op}(c)(t)\mathcal{F}^{-1}= \int^{\oplus}_{\rr}\Op(c(t, \lambda))d\lambda, \ c(t, \lambda)\in \cinfb(\rr; \widetilde{S}^{m}(T^{*}\Sigma)),\\[2mm]
 &w\in \cinfb(\rr; \cW^{-\infty}(\Sigma\times \rr)), \ [w(t), T_{l}]=0\\[2mm]
\Leftrightarrow& \ \mathcal{F}w(t)\mathcal{F}^{-1}= \int^{\oplus}_{\rr}w(t, \lambda)d\lambda, \ w(t, \lambda)\in \cinfb(\rr; \widetilde{\cW}^{-\infty}(\Sigma)). 
 \eea
 \end{equation}
  The same results hold also if we replace $\cinfb(\rr; A)$ by $S^{\delta}(\rr; A)$ for $A= S^{m}_{\rm ph}(T^{*}(\Sigma\times \rr))$, $\widetilde{S}^{m}(T^{*}\Sigma)$ etc. In fact it suffices to note that $c(t)\in S^{\delta}(\rr; A)$ iff $\langle t\rangle^{-\delta+n}\p_{t}^{n}c(t)\in \cinfb(\rr; A)$ for all $n\in \nn$.

{\it Step 3}. Let now $a(t)\in S^{0}(\rr; \Psi^{2}(\Sigma))$  be as in the lemma and let $A(t)= a(t)+ D_{l}^{2}$ acting on $L^{2}(\Sigma\times \rr)$. The operator $A(t)$ is elliptic in $S^{0}(\rr; \Psi^{2}(\Sigma\times \rr))$, selfadjoint on $H^{2}(\Sigma\times \rr)$ with $A(t)\geq c_{0}\one$ for $c_{0}$ as in the lemma. 

 We would like to apply Thm. \ref{seeley}  for $\alpha= -1$ to the class $S^{0}(\rr; \Psi^{m}(\Sigma\times \rr))$ to  obtain that $A(t)^{-1}$ belongs to $S^{0}(\rr; \Psi^{-2}(\Sigma\times \rr))$.
Note that  the ideal $S^{0}(\rr; \cW^{-\infty}(\Sigma\times \rr))$ is smaller than the one used in 
Subsect. \ref{pdosec} for the manifold $\Sigma\times \rr$, because  multi-commutators ${\rm ad}^{\gamma}_{l}A$ appear in \eqref{comutoto}.

However we can still apply the abstract framework in \cite{alnv1} to this situation, see \cite[Subsect. 5.3.4]{bounded} for a concise summary.  We choose as Hilbert space $\cH= L^{2}(\rr_{t}\times \Sigma_{\rx}\times \rr_{l})$.   As injective operator in $S^{0}(\rr; \cW^{-\infty}(\Sigma\times \rr))$ we choose $\e^{\Delta_{h}+ D_{l}^{2}+1}$. The only point which differs a little from the situation in Subsect. \ref{pdosec} is the {\em spectral invariance} of $S^{0}(\rr; \cW^{-\infty}(\Sigma\times \rr))$, see the proof of Thm. \ref{seeley}: if $R\in S^{0}(\rr; \cW^{-\infty}(\Sigma\times \rr))$ and $(\one +R)$ is boundedly invertible in $B(\cH)$, then we have
\[
(\one +R)^{-1}= \one+ R_{1}, \ R_{1}= - R + R(\one + R)^{-1}R.
\]
We have $R_{1}= \int^{\oplus}_{\rr}R_{-1}(t)dt$, for 
\beq\label{titata}
R_{1}(t)= - R(t) + R(t)(\one + R(t))^{-1}R(t).
\eeq 
 We have to check that  \[
 \|\p_{t}^{n}{\rm ad}_{l}^{\gamma}R_{-1}(t)\|_{B(H^{-p}(\Sigma\times \rr), H^{p}(\Sigma\times \rr))}\in O(\bra t\ket^{-n}),\ \forall \gamma, n,p\in \nn.
\]
This follows from \eqref{titata} using the Leibniz rule for $\p_{t}$ and ${\rm ad}_{l}$ and the identities
\[
\begin{array}{l}
\p_{t}(\one+R(t))^{-1}= - (\one + R(t))^{-1}\p_{t}R(t)(\one + R(t))^{-1}\\[2mm]
{\rm ad}_{l}(\one+R(t))^{-1}= - (\one + R(t))^{-1}{\rm ad}_{l}R(t)(\one + R(t))^{-1}.
\end{array}
\]
In conclusion we can apply Seeley's theorem and obtain that $A(t)^{-1}$ belongs to $S^{0}(\rr; \Psi^{-m}(\Sigma\times \rr))$. We have then:
\[
\mathcal{F}A(t)^{-1}\mathcal{F}^{-1}= \int^{\oplus}_{\rr}(a(t)+ \lambda^{2})^{-1}d\lambda, 
\]
which by \eqref{e.deco3} implies that $(a(t)+ \lambda^{2})^{-1}\in S^{0}(\rr; \widetilde{\Psi}^{-2}(\Sigma))$.  This concludes the proof of the lemma.\qed

{\bf \noindent Proof of Prop. \ref{l5.1}.} In view of the identity
\[
a_{1}^{1+ \alpha}- a_{2}^{1+ \alpha}= (a_{1}- a_{2})a_{1}^{\alpha}+ a_{2}(a_{1}^{\alpha}- a_{2}^{\alpha}),
\]
 we see that it suffices to prove the proposition for $0<\alpha<1$. We  will use the following formula, valid for example  if $a$ is a selfadjoint operator on a Hilbert space $\cH$  with $a\geq c\one$, $c>0$:
\beq\label{powers}
a^{\alpha}= C_{\alpha}\int_{0}^{+\infty}(a+ s)^{-1}s^{\alpha}ds= C_{\alpha}\int_{\rr}(a+ \lambda^{2})^{-1}\lambda^{2\alpha+1}d\lambda, \ \alpha\in \rr,
\eeq
where the integrals are norm convergent in say, $B(\Dom a^{m}, \cH)$ for $m$ large enough.

We have for $r(t)= a_{1}(t)- a_{2}(t)$:
\[
\bea
(a_{1}(t)+ \lambda^{2})^{-1}&= ( a_{2}(t)+ \lambda^{2})^{-1}(\one + r(t)(a_{1}(t)+ \lambda^{2})^{-1})\\
&= ( a_{2}(t)+ \lambda^{2})^{-1}+  ( a_{2}(t)+ \lambda^{2})^{-2}(a_{2}(t)+ \lambda^{2})r(t)(a_{1}(t)+ \lambda^{2})^{-1}\\
&= (a_{2}(t)+ \lambda^{2})^{-1}+ (a_{2}(t)+ \lambda^{2})^{-2}a_{2}(t)c_{1}(t, \lambda)\\
&= (a_{2}(t)+ \lambda^{2})^{-1}+ a_{2}(t)c_{2}(t, \lambda), 
\eea
\]
where using Lemma \ref{lemomo}, $c_{1}(t,\lambda)\in S^{-\delta}(\rr; \widetilde{\Psi}^{0}(\Sigma))$ and $c_{2}(t, \lambda)\in S^{-\delta}(\rr; \widetilde{\Psi}^{-4}(\Sigma))$. From \eqref{powers} we obtain that:
\beq\label{potopoto}
a_{1}^{\alpha}(t)- a_{2}^{\alpha}(t)= C_{\alpha}a_{2}(t)\int_{\rr}c_{2}(t, \lambda)\lambda^{2\alpha+1}d\lambda.
\eeq
We now write $c_{2}(t, \lambda)$ as $\Op(d_{2}(t, \lambda))+ w_{2}(t, \lambda)$, for $d_{2}\in S^{-\delta}(\rr; \widetilde{S}_{\rm ph}^{-4}(T^{*}\Sigma))$ and $w_{2}(t, \lambda)\in S^{-\delta}(\rr; \widetilde{\cW}^{-\infty}(\Sigma))$.
Using that
\[
\int_{\rr}(\langle \xi\rangle + \langle \lambda \rangle)^{-4-k}\lambda^{2\alpha+1}d\lambda\sim \langle \xi\rangle^{2\alpha-2-k},
\]
we first obtain that
\[
\int_{\rr}d_{2}(t, \lambda)\lambda^{2\alpha+1}d\lambda\in S^{-\delta}(\rr; S_{\rm ph}^{2\alpha-2}(\Sigma)).
\]
Similarly we obtain that $\int_{\rr}w_{2}(t, \lambda)\lambda^{2\alpha+1}d\lambda\in S^{-\delta}(\rr; \cW^{-\infty}(\Sigma))$.  Using \eqref{potopoto} this implies that $a_{1}^{\alpha}(t)- a_{2}^{\alpha}(t)\in S^{-\delta}(\rr; \Psi^{2\alpha}(\Sigma))$, as claimed.
\subsection{Proof of Prop. \ref{p5.1}}\label{to1}

We follow the proof in \cite{GW}. The ${\rm out}$ and ${\rm in}$ cases are treated similarly.
We set $a_{0}= \frac{\i}{2}(\epsilon^{-1}\p_{t}\epsilon+ \epsilon^{-1}r\epsilon)$,
\[
F(c)\defeq  \12 \epsilon^{-1}\left( \i \p_{t}c+ [\epsilon, c]+ \i rc - c^{2}\right)= G(c)-\12 \epsilon^{-1}c^{2}.
\]
and look for $b(t)$ as $\epsilon(t)+ c$, where  $c\in \Psi_{\td}^{0, -1- \delta}$ has to satisfy $c= a_{0}+ F(c)$.
Let us start by studying some properties of  the map $F$. First if $c\in \Psi^{0, -\mu}_{\td}$ then:
\[
\begin{array}{l}
G(c)\in \Psi_{\td}^{-1, 0}\times \Psi_{\td}^{0, -1- \mu}+ \Psi_{\td}^{-1, 0}\times \Psi_{\td}^{0, -\mu}+ \Psi_{\td}^{-1, 0}\Psi_{\td}^{0, -1- \delta}\times \Psi_{\td}^{0, -\mu},\\[2mm]
\epsilon^{-1}c^{2}\in \Psi_{\td}^{-1, -2\mu},
\end{array}
\]
hence 
\begin{equation}
\label{tito}
c\in \Psi_{\td}^{0, -\mu}\ \Rightarrow \ F(c)\in \Psi_{\td}^{-1, -\mu}.
\end{equation}
Secondly, if $c_{1}, c_{2}\in \Psi_{\td}^{0, -\mu}$ and $c_{1}- c_{2}\in \Psi_{\td}^{-j, -\mu}$ then:
\[
\bea
&G(c_{1})- G(c_{2})= G(c_{1}-c_{2})\\[1mm]
&\in \Psi_{\td}^{-1, 0}\times \Psi_{\td}^{-j, -1-\mu}+ \Psi_{\td}^{-1, 0}\times\Psi_{\td}^{-j, -\mu}+ \Psi_{\td}^{-1, 0}\times \Psi_{\td}^{0, -1- \delta}\times \Psi_{\td}^{-j, -\mu},\\[1mm]
&\epsilon^{-1}(c_{1}^{2}- c_{2}^{2})
=\epsilon^{-1}c_{1}(c_{1}- c_{2})+ \epsilon^{-1}(c_{1}- c_{2})c_{2}\\[1mm]
&\in \Psi_{\td}^{-1, 0}\times \Psi_{\td}^{0, -\mu}\times \Psi_{\td}^{-j, -\mu}+ \Psi_{\td}^{-1, 0}\times \Psi_{\td}^{-j, -\mu}\times \Psi_{\td}^{0, -\mu},
\eea
\] 
hence
\begin{equation}
\label{titu}
c_{1}, c_{2}\in  \Psi_{\td}^{0, -\mu}, \ c_{1}- c_{2}\in \Psi_{\td}^{-j, -\mu}\ \Rightarrow \ F(c_{1})- F(c_{2})\in\Psi_{\td}^{-j-1, -\mu}.
\end{equation}
 We also have
 \[
 \bea
&a_{0}= \frac{\i}{2}(\epsilon^{-1}\p_{t}\epsilon+ \epsilon^{-1}r\epsilon)\\[1mm]
&\quad\in  \Psi_{\td}^{-1, 0}\times  \Psi_{\td}^{1, -1-\delta}+  \Psi_{\td}^{-1, 0}\times  \Psi_{\td}^{0, -1- \delta}\times \Psi_{\td}^{1,0}\subset \Psi_{\td}^{0, -1- \delta}.
\eea
\]
We now follow the proof in \cite[Lemma A.1]{GW}, namely, we set $c_{0}= a_{0}$, $c_{n}= a_{0}+ F(c_{n-1})$ and obtain by induction that 
$c_{n}- c_{n-1}\in \Psi_{\td}^{-n, -1-\delta}$.
 We then set 
 \[
c\sim a_{0}+ \sum_{n=1}^{\infty}c_{n}- c_{n-1}\in \Psi_{\td}^{0, -1-\delta},
 \]
where the $\Psi_{\td}^{0, -1-\delta}$ membership follows from Lemma \ref{l5.2}. We obtain that
 \[
\i\p_{t}b- b^{2}+ a + \i rb\in \Psi_{\td}^{-\infty,-1-\delta}.
\]
 By construction we have $b(t)= \epsilon(t)+ \Psi_{\td}^{0, -1-\delta}(\rr; \Sigma)$. Applying Prop. \ref{l5.1} we get
 \[
\epsilon(t)= \epsilon_{\outin}+ \Psi_{\td}^{1, -\delta}(\rr; \Sigma) \hbox{ on }\rr^{\pm}\times \Sigma.
\]
 \qed
 \subsection{Proof of Prop. \ref{propoesti}}\label{to2}
 From Prop. \ref{p5.1} we first obtain that $b^{+}- b^{-}= (b+ b^{*})= 2\epsilon+ \Psi^{0, -1- \delta}_{\td}(\rr; \Sigma)$. It follows first that $(b^{+}- b^{-})^{2}= 4 a+ \Psi_{\td}^{1, -1- \delta}(\rr; \Sigma)$ and then by Prop. \ref{p5.1} that
\[
(b^{+}- b^{-})^{\alpha}= ((b^{+}- b^{-})^{2})^{\alpha/2}= \begin{cases}
(2 \epsilon)^{\12}+ \Psi^{0, - 1- \delta}_{\td}(\rr; \Sigma), \ \alpha= \12\\
(2\epsilon)^{-\12}+ \Psi^{-3/2, -1- \delta}_{\td}(\rr; \Sigma), \ \alpha= -\12.
\end{cases}
\]
We obtain again by Prop. \ref{p5.1} that:
\[
\begin{array}{l}
[(b^{+}- b^{-})^{-\12}, b^{\pm}]= [(2\epsilon)^{-\12}+ \Psi_{\td}^{-3/2, -1- \delta}(\rr; \Sigma), \pm \epsilon+ \Psi^{0, -1- \delta}_{\td}(\rr; \Sigma)] \\ \phantom{[(b^{+}- b^{-})^{-\12}, b^{\pm}]}\in \Psi^{-3/2, -1- \delta}_{\td}(\rr; \Sigma),\\[2mm]
\p_{t}(b^{+}- b^{-})^{-\12}(b^{+}- b^{-})^{\12}= (\p_{t}(2\epsilon)^{-\12}+ \Psi^{-3/2, -2- \delta}_{\td}(\rr; \Sigma))\times \Psi^{\12, 0}_{\td}(\rr; \Sigma)\\[2mm]
=\p_{t}(2\epsilon)^{-\12}\times\Psi^{\12, 0}_{\td}(\rr; \Sigma)+ \Psi^{-1, -2-\delta}_{\td}(\rr; \Sigma).
\end{array}
\]
Since by Prop. \ref{l5.1} $(2\epsilon)^{-\12}= (2 \epsilon_{\outin})^{-\12}+ \Psi^{-3/2, -\delta}_{\td}(\rr^{\pm}; \Sigma)$, we have 
\[
\p_{t}(2\epsilon)^{-\12}\in \Psi^{-3/2, -1- \delta}_{\td}(\rr; \Sigma)\ \Rightarrow \ \p_{t}(b^{+}- b^{-})^{-\12}(b^{+}- b^{-})^{\12}\in \Psi^{-1, -1- \delta}_{\td}(\rr; \Sigma).
\]  Since by hypothesis $(\td)$, $r\in \Psi^{0, -1- \delta}_{\td}(\rr; \Sigma)$, we obtain that $r_{b}^{\pm}\in \Psi^{0, -1- \delta}_{\td}(\rr; \Sigma)$. Finally we obtain immediately from Prop. \ref{p5.1}  that $r_{-\infty}^{\pm}= \i\p_{t}b^{\pm}- (b^{\pm})^{2}+ a + \i rb^{\pm}\in \Psi^{-\infty, -1- \delta}_{\td}(\rr; \Sigma)$. \qed

\subsection{Proof of Lemma \ref{l10.1}}\label{apoti}
Let us fix two good chart coverings $\{U_{i}, \psi_{i}\}_{i\in \nn}$ and $\{\tilde{U}_{i}, \tilde{\psi}_{i}\}_{i\in \nn}$ with $U_{i}\Subset \tilde{U}_{i}$.
Since $\altb\in \cinfb(\rr; \BT^{1}_{0}(\Sigma, \altk))$, we obtain easily by transporting $\altb$ to $B_{n}(0,1)$ using $\psi_{i}$ that there exists $t_{+,\epsilon}>0$ such that $\ry(t,s, \cdot)$ is a bounded diffeomorphism of $(\Sigma, \altk)$, uniformly  for $|t-s|\leq t_{+,\epsilon}$. By the group property  of the flow we can replace $t_{+,\epsilon}$ by any $\varT>0$, keeping the above uniformity property.

Moreover if $\altb_{i}\defeq  (\psi_{i}^{-1})^{*}b$ we obtain from $(\ast)$ that $\altb_{i}\in S^{-\delta}(\rr; \BT^{1}_{0}(B_{n}(0,1)))$, uniformly in $i\in \nn$. If $\ry_{i}(t, s, \cdot)$ denotes the flow of $\altb_{i}$ we obtain that:
\[
\ry_{i}(t,s, \rx)= \rx+ \int_{s}^{t}\altb_{i}(\sigma, \ry_{i}(\sigma, s, \rx))d\sigma.
\]
From this we obtain that there exists $\varT\gg 1$ such that \[
\ry_{i}(\pm t, \pm \varT, \cdot): B_{n}(0, \textstyle\12)\to B_{n}(0,1)
\] for all $t\geq \varT$ and moreover 
\[
\lim_{t\to \pm \infty}\ry_{i}(t, \pm \varT, y)= \int_{\pm \varT}^{\pm \infty}\altb_{i}(\sigma, \ry_{i}(\sigma, \pm \varT, \rx))ds\eqdef \ry_{i}(\pm\infty, \varT, \rx).
\]
We can also choose $\varT$ large enough so that if we set
\beq\label{e10.3}
\ry(\pm\infty, \pm \varT,\rx)\defeq  \psi_{i}^{-1}\circ \ry_{i}(\pm\infty, \pm \varT, \cdot)\circ \psi_{i}(y), \ \rx\in U_{i}
\eeq
then $\ry(\pm\infty, \pm \varT, \cdot)$ is well defined, and is a bounded diffeomorphism of $(\Sigma, \altk)$.  We now set:
\[
\ry_{\outin}\defeq  \ry(\pm\infty, \pm \varT, \cdot)\circ \ry(\pm \varT, 0, \cdot),
\]
which is also a bounded diffeomorphism of $(\Sigma, \altk)$. We also obtain from \eqref{e10.3} and the previous estimates on  $\ry(t, s, \cdot)$ for $|t-s|\leq \varT$ that $\{\ry(t,0,  \cdot )\}_{t\in \rr}$ is a bounded family of bounded diffeomorphisms of $(\Sigma, \altk)$.
Moreover from \eqref{e10.3} we obtain that
\beq\label{e10.4}
\ry_{i}(t, 0, \rx)-\ry_{i, \outin}(\rx)\in S^{1- \delta'}(\rr; \cinfb(B_{n}(0,1))), \hbox{ uniformly in }i\in \nn.
\eeq
Let us now consider the metric $\chi^{*}\altg$.

Since $v\cdot dt=0$, $\chi^{*}\altg= {}^{t}\!D\chi(\altg\circ \chi)D\chi= - \altch^{2}(t, \rx)dt^{2}+ \hat \alth(t, \rx)d\rx^{2}$. Using  \eqref{e10.4} we obtain that
\[
\bea
\altch(t, \rx)&= \altc(t, \ry(t, \rx))+ S^{-2\delta'}(\rr; \BT^{0}_{0}(\Sigma))\\[2mm]
&= \altc_{\outin}(\ry(t,\rx))+ S^{-\min(2\delta', \delta)}(\rr^{\pm}; \BT^{0}_{0}(\Sigma))\\[2mm]
&=\altc_{\outin}(\ry_{\outin}(\rx))+ S^{-\min(1- \delta', \delta)}(\rr^{\pm}; \BT^{0}_{0}(\Sigma)).
\eea
\]
Similarly,
\[
\bea
\hat \alth(t, \rx)&= {}^{t}\!D\ry(t, \rx)\alth(t, \ry(t, \rx))D\ry(t, \rx)\\[2mm]
&= {}^{t}\!D\ry(t, \rx)\alth_{\outin}(\ry(t, \rx))D\ry(t, \rx)+ S^{-\delta}(\rr^{\pm}; \BT^{0}_{2}(\Sigma))\\[2mm]
&= {}^{t}\!D\ry_{\outin}(\rx)\alth_{\outin}(\ry_{\outin}(\rx))D\ry_{\outin}(\rx) +S^{-\min(1- \delta', \delta)}(\rr^{\pm}; \BT^{0}_{2}(\Sigma)),\\[2mm]
\chi^{*}\altV&=r(t,\ry(t, \rx))= r_{\outin}(\ry(t, \rx))+ S^{-\delta}(\rr^{\pm}; \BT^{0}_{0}(\Sigma))\\[2mm]
&=\altV_{\outin}(\ry_{\outin}(\rx))+  S^{-\min(1- \delta', \delta)}(\rr^{\pm}; \BT^{0}_{0}(\Sigma)).
\eea
\]
Since by definition
\[
 \hat \alth_{\outin}= \ry_{\outin}^{*}\alth_{\outin},\  \ \altch_{\outin}= \ry_{\outin}^{*}\altc_{\outin}, \ \ \altVh_{\outin}= \ry_{\outin}^{*}\altV_{\outin},
\]
we obtain the assertion. \qeds

\subsection*{Acknowledgments} The authors are grateful to Jan Derezi\'nski and Andr\'as Vasy for stimulating discussions. M.\,W. gratefully acknowledges the France-Stanford Center for Interdisciplinary Studies for financial support and the Department of Mathematics of Stanford University, where part of the work was performed. M.\,W. gratefully acknowledges partial support from the National Science Center, Poland, under the grant UMO-2014/15/B/ST1/00126. The authors also wish to thank the Erwin Schr\"odinger Institute in Vienna for its hospitality during the program ``Modern theory of wave equations''.

\end{document}